%% file: Main.tex
\documentclass[acmsmall]{acmart}

\usepackage{colortbl}
\usepackage{multirow}
\usepackage{tabularx}
\usepackage{tcolorbox}
\usepackage[toc]{appendix}
\usepackage{longtable}
\usepackage{supertabular}
\usepackage{tabularray}
\usepackage{tabularx}
\usepackage[T1]{fontenc}
\usepackage{color, colortbl}
\usepackage{booktabs}
\usepackage{xcolor}
\usepackage{colortbl}
\definecolor{LightGray}{gray}{0.9}
\usepackage{tikz}
\usepackage[normalem]{ulem}
\usepackage{fontawesome5}
\usepackage{soul}
\usepackage{tikz}
\usepackage{array}
\usepackage{booktabs}
\usepackage{natbib}
\definecolor{darkgreen}{rgb}{0.0, 0.5, 0.0}
\definecolor{lightgreen}{HTML}{95d5b2}

\definecolor{lightred}{rgb}{1.0, 0.8, 0.8}
\definecolor{lightgreen}{rgb}{0.8, 1.0, 0.8}
\definecolor{darkgreen}{rgb}{0.6, 0.8, 0.6} % slightly darker than lightgreen
\definecolor{neutralgray}{gray}{0.8}

\newcommand{\fivepointlikert}[5]{%
  \begin{tikzpicture}[baseline]
    \pgfmathsetmacro{\barwidth}{4.0} % Adjust width as needed

    \pgfmathsetmacro{\widthSD}{#1*\barwidth/100}
    \pgfmathsetmacro{\widthD}{#2*\barwidth/100}
    \pgfmathsetmacro{\widthN}{#3*\barwidth/100}
    \pgfmathsetmacro{\widthA}{#4*\barwidth/100}
    \pgfmathsetmacro{\widthSA}{#5*\barwidth/100}

    \fill[darkgreen] (0,0) rectangle (\widthSD,0.5);
    \fill[lightgreen] (\widthSD,0) rectangle (\widthSD+\widthD,0.5);
    \fill[neutralgray] (\widthSD+\widthD,0)
                      rectangle (\widthSD+\widthD+\widthN,0.5);
    \fill[lightred] (\widthSD+\widthD+\widthN,0)
                     rectangle (\widthSD+\widthD+\widthN+\widthA,0.5);
    \fill[red] (\widthSD+\widthD+\widthN+\widthA,0)
                    rectangle (\widthSD+\widthD+\widthN+\widthA+\widthSA,0.5);

    \draw[black] (0,0) rectangle (\barwidth,0.5);

    \ifnum#1>0
      \node[font=\small, text=white] at (\widthSD/2, 0.25) {#1\%}; % SD
    \fi
    \ifnum#2>0
      \node[font=\small, text=black]
        at (\widthSD + \widthD/2, 0.25) {#2\%}; % D
    \fi
    \ifnum#3>0
      \node[font=\small, text=black]
        at (\widthSD + \widthD + \widthN/2, 0.25) {#3\%}; % N
    \fi
    \ifnum#4>0
      \node[font=\small, text=black]
        at (\widthSD + \widthD + \widthN + \widthA/2, 0.25) {#4\%}; % A
    \fi
    \ifnum#5>0
      \node[font=\small, text=black]
        at (\widthSD + \widthD + \widthN + \widthA + \widthSA/2, 0.25) {#5\%}; % SA
    \fi
  \end{tikzpicture}%
}

\newcommand{\fivepointlikertt}[5]{%
  \begin{tikzpicture}[baseline]
    \pgfmathsetmacro{\barwidth}{4.0} % Adjust width as needed
    \pgfmathsetmacro{\totalsample}{8} % Total sample size

    \pgfmathsetmacro{\widthSD}{#1*\barwidth/\totalsample}
    \pgfmathsetmacro{\widthD}{#2*\barwidth/\totalsample}
    \pgfmathsetmacro{\widthN}{#3*\barwidth/\totalsample}
    \pgfmathsetmacro{\widthA}{#4*\barwidth/\totalsample}
    \pgfmathsetmacro{\widthSA}{#5*\barwidth/\totalsample}

    \fill[darkgreen] (0,0) rectangle (\widthSD,0.5); % Strongly Disagree
    \fill[lightgreen] (\widthSD,0) rectangle (\widthSD+\widthD,0.5); % Disagree
    \fill[neutralgray] (\widthSD+\widthD,0) rectangle (\widthSD+\widthD+\widthN,0.5); % Neutral
    \fill[lightred] (\widthSD+\widthD+\widthN,0) rectangle (\widthSD+\widthD+\widthN+\widthA,0.5); % Agree
    \fill[red] (\widthSD+\widthD+\widthN+\widthA,0) rectangle (\widthSD+\widthD+\widthN+\widthA+\widthSA,0.5); % Strongly Agree

    \draw[black] (0,0) rectangle (\barwidth,0.5);

    \ifnum#1>0
      \node[font=\small, text=white] at (\widthSD/2, 0.25) {#1}; % SD
    \fi
    \ifnum#2>0
      \node[font=\small, text=black] at (\widthSD + \widthD/2, 0.25) {#2}; % D
    \fi
    \ifnum#3>0
      \node[font=\small, text=black] at (\widthSD + \widthD + \widthN/2, 0.25) {#3}; % N
    \fi
    \ifnum#4>0
      \node[font=\small, text=black] at (\widthSD + \widthD + \widthN + \widthA/2, 0.25) {#4}; % A
    \fi
    \ifnum#5>0
      \node[font=\small, text=black] at (\widthSD + \widthD + \widthN + \widthA + \widthSA/2, 0.25) {#5}; % SA
    \fi
  \end{tikzpicture}%
}

\tcbset{
    mybox/.style={
        colback=gray!10, % Background color
        colframe=black, % Border color
        width=\linewidth, % Flexible width
        boxrule=1mm, % Thickness of the left border
        leftrule=1mm, % Extra thick left rule
        rightrule=0mm, % No right rule
        toprule=0mm, % No top rule
        bottomrule=0mm, % No bottom rule
        sharp corners, % Square corners
        boxsep=0pt, % No extra padding
        left=5pt, % Padding on the left
        right=10pt, % Padding on the right
        top=5pt, % Padding on the top
        bottom=5pt % Padding on the bottom
    }
}

\definecolor{labelColorBig}{RGB}{99,180,255}

\newcommand{\overviewLabelBig}[1]{%
  \tikz[baseline={(current bounding box.center)}]%
    \node[fill=labelColorBig, text=white, circle, inner sep=0.05em, minimum size=1.5em] {#1};%
}

\newif\ifdraft
\draftfalse %hides boldifications
\newcommand{\new}[1]{\textcolor{black!60!black}{#1}}
\def\checkmark{\tikz\fill[scale=0.4](0,.35) -- (.25,0) -- (1,.7) -- (.25,.15) -- cycle;}
\usepackage[framemethod=tikz]{mdframed} % 
\def\BibTeX{{\rm B\kern-.05em{\sc i\kern-.025em b}\kern-.08em
    T\kern-.1667em\lower.7ex\hbox{E}\kern-.125emX}}

\AtBeginDocument{%
  \providecommand\BibTeX{{%
    \normalfont B\kern-0.5em{\scshape i\kern-0.25em b}\kern-0.8em\TeX}}}

\setcopyright{cc}
\setcctype{by}
\acmJournal{TOSEM}
\acmYear{2025} \acmVolume{1} \acmNumber{1} \acmArticle{1} \acmMonth{1} \acmPrice{}\acmDOI{10.1145/3769012}


\begin{document}

\title{\new{Addressing OSS Community Managers' Challenges in Contributor Retention}}

\author{Zixuan Feng}
\email{fengzi@oregonstate.edu}
\orcid{0000-0001-9163-6853}
\affiliation{%
  \institution{Oregon State University}
  \city{Corvallis}
  \state{Oregon}
  \country{USA}
}

\author{Katie Kimura}
\orcid{0009-0009-3858-0905}
\affiliation{%
  \institution{Oregon State University}
  \city{Corvallis}
  \state{Oregon}
  \country{USA}}
\email{kimuraka@oregonstate.edu}

\author{Bianca Trinkenreich}
\orcid{0000-0001-7302-6082}
\affiliation{%
  \institution{Colorado State University}
  \city{Fort Collins}
   \state{Colorado}
  \country{USA}}
\email{bianca.trinkenreich@colostate.edu}

\author{Igor Steinmacher}
\orcid{0000-0002-0612-5790}
\affiliation{%
  \institution{Northern Arizona University}
  \city{Flagstaff}
   \state{Arizona}
  \country{USA}}
\email{Igor.Steinmacher@nau.edu}

\author{Marco Gerosa}
\orcid{0000-0003-1399-7535}
\affiliation{%
  \institution{Northern Arizona University}
  \city{Flagstaff}
   \state{Arizona}
  \country{USA}}
\email{Marco.Gerosa@nau.edu}

\author{Anita Sarma}
\orcid{0000-0002-1859-1692}
\affiliation{%
  \institution{Oregon State University}
  \city{Corvallis}
  \country{USA}}
\email{Anita.Sarma@oregonstate.edu}

\begin{abstract}

Open-source software (OSS) community managers face significant challenges in retaining contributors, as they must monitor activity and engagement while navigating complex dynamics of collaboration. Current tools designed for managing contributor retention (e.g., dashboards) fall short by providing retrospective rather than predictive insights to identify potential disengagement early. Without understanding how to anticipate and prevent disengagement, new solutions risk burdening community managers rather than supporting retention management. Following the Design Science Research paradigm, we employed a mixed-methods approach for problem identification and solution design to address contributor retention. To identify the challenges hindering retention management in OSS, we conducted semi-structured interviews, a multi-vocal literature review, and community surveys. Then through an iterative build-evaluate cycle, we developed and refined strategies for diagnosing retention risks and informing engagement efforts. We operationalized these strategies into a web-based prototype, incorporating feedback from 100+ OSS practitioners, and conducted an in situ evaluation across two OSS communities. 
Our study offers (1) empirical insights into the challenges of contributor retention management in OSS, (2) actionable strategies that support OSS community managers’ retention efforts, and (3) a practical framework for future research in developing or validating theories about OSS sustainability.

\end{abstract}


\maketitle

\input{section/sec1_Intro}
\label{sec:intro}

\input{section/sec2_RW}

\label{sec:rw}

\input{section/sec3_overall_method}

\label{sec:overall_method}

\input{section/sec4_challenge}

\label{sec:c_s}

\input{section/sec5_stratagies}

\input{section/sec6_FGD}

\input{section/sec7_User_evaluation}

\input{section/sec8_Discussion}

\label{sec:discussion}

\input{section/sec9_Threats}

\label{sec:threats}

\input{section/sec10_Conclusion}

\label{sec:conclusion}

\bibliographystyle{ACM-Reference-Format}
\bibliography{ref, gray}

% % Gray refs section
% \begin{refsection}
% \nocite{G1,G2,G3,G4,G5,G6,G7,G8,G9,G10,%
% G11,G12,G13,G14,G15,G16,G17,G18,G19,G20,%
% G21,G22,G23,G24,G25,G26,G27,G28,G29,G30,%
% G31,G32,G33,G34,G35}
% \bibliographystyle{ACM-Reference-Format}
% \bibliography{gray}
% \end{refsection}

% % Normal refs
% % \begin{refsection}
% \bibliographystyle{ACM-Reference-Format}
% \bibliography{ref}
% % \end{refsection}

% \clearpage
% \newpage

% \nocitesec{*}
% \bibliographystylesec{appendixStyle}
% \bibliographysec{gray}

\end{document}

%% file: section/sec1_Intro.tex
\section{Introduction}

Managing contributor retention in Open Source Software (OSS) projects is inherently challenging due to their volunteer-driven, peer-based production model, which lacks formal organizational structures and stable commitments. These challenges are further amplified by the volatile relationship companies have with OSS, as corporate priorities shift and funding fluctuates \cite{qiu2019going, zhou2012make, dabbish2012social}. OSS communities need active and sustained participation to remain healthy \cite{alfayez2017does}. High turnover rates in OSS hinder project progress, threaten the sustainability of the community, and risk project failure \cite{zhang2022turnover, zhang2022corporate, zhou2012make, zhang2019companies, vasilescu2015gender}. Without practical tools to anticipate and mitigate these risks, unexpected contribution breaks may disrupt the sustainability of projects~\cite{iaffaldano2019developers}.
When contributors leave due to burnout, lack of recognition, shifting personal interests and priorities, or difficulty integrating within the community, projects risk losing skill sets and institutional knowledge \cite{trinkenreich2020hidden}.

Even though project maintainers recognize the importance of retaining contributors, they acknowledge its difficulty, as one participant from our study noted, \textit{``It's hard to get more contributors engaged in [the project]''} [P1] \footnote{Interview participant ID.}, and another commenting, \textit{``We've had some people who have been very committed to this project, [but] definitely have burnt out of contributing to it''} [P4]. Recognizing the significance of this challenge, existing research has proposed a range of interventions. For instance, \citet{guizani2022attracting} introduced a prototype to support managers in attracting and retaining newcomers, while \citet{qiu2023climate} and \citet{ramchandran2022exploring} developed dashboards that help identify community health trends and highlight activity downturn events. Industry tools like DevStats \cite{cncfDevStats} and the Amazon OSS Dashboard \cite{amzn_oss_dashboard} provide static snapshots of contributors’ activities. Although these visualizations offer valuable data, they are often limited to descriptive analytics, which help projects mitigate issues after the fact and are not proactive. Another line of research draws inspiration from health-related studies \cite{lee1997survival, guo2010survival}, predicting contributor retention based on historical contribution patterns \cite{bao2019large, eluri2021predicting, giovanini2021leveraging, calefato2022will}. However, these models have yet to be integrated into tools for day-to-day community management. Without fully understanding community managers' challenges, continuously introducing new interventions may not effectively support community managers. Instead, identifying these challenges should be the first step in designing meaningful support mechanisms.

Thus, while several works aim to tackle contributor retention, community managers still struggle with where, when, and how to intervene to retain their contributors \cite{guizani2022attracting, gray2022disengage}. In this work, we close this gap by providing a deeper understanding of both the challenges managers face and the strategies that could support their retention efforts, offering insights into where interventions are most needed, when community managers should act, and how different strategies can be implemented to sustain engagement. Without such an understanding, well-intentioned research-driven interventions risk adding more complexity rather than alleviating it. Our work is guided by the following two research questions:

\begin{tcolorbox}[mybox] 

\noindent \textbf{RQ1:} What are the challenges in proactively managing retention in OSS?  

\textbf{RQ2:} \new{What strategies can support OSS community managers in diagnosing and managing retention challenges?}

\end{tcolorbox}

To answer these questions, we followed the Design Science Research (DSR) paradigm \cite{runeson2020design, engstrom2020software}, which aims to address real-world software engineering challenges by creating and evaluating artifacts that generate actionable design knowledge. We conceptualized the problem of contributor retention in OSS as a recurring and socio-technical issue that lacks operationalizable, community-centered solutions. We first identified 10 core challenges in contributor retention management through a triangulated empirical analysis involving interviews with OSS community managers (\faComments), multi-vocal literature review (academic and industry/gray literature \cite{}) (\faBook), and expert validation surveys (\faPollH).

In response to these empirically grounded challenges, we then developed actionable strategies that aim to empower community managers with practical insights for supporting retention management. We compiled a list of nine strategies based on insights from the literature review (\faBook). These strategies were operationalized through a prototype using an iterative build–evaluate cycle and a user-centric design process co-developed with input from over 100 OSS practitioners (\faUsers).  The prototype served both as an intervention and a validation scaffold to ensure the strategies' relevance and usability in practice.

To validate our intervention and move toward technological rule generalization, we evaluated our prototype in two distinct OSS communities—Pyomo \cite{pyomo2024} and DeepSpeed \cite{deepspeed2024} (\faUserEdit)—each with different governance structures and community cultures. We loaded their project data into the prototype and invited their governance teams to evaluate the prototype. This approach ensured our strategies were both theoretically grounded and practically applicable to OSS projects.

Through this study, we aligned research advancements with the pragmatic needs of practitioners by exploring the realities of retention management. We also proposed integrated strategies to help OSS communities engage and retain contributors proactively. Our study provides (1) empirical insights into the challenges OSS community managers face in sustaining contributor engagement, along with a structured analysis of the strategies that support retention efforts; (2) a human-factors-driven framework that serves as a replicable blueprint for future researchers seeking to develop or validate theories about OSS sustainability and software engineering processes; and (3) practical guidance for OSS practitioners, offering diagnostic strategies to surface disengagement risks and inform retention-related decisions.

%% file: section/sec2_RW.tex
\section{Background and Related Work}

Effective retention management is vital for OSS project sustainability, as it preserves critical expertise, prevents contributor attrition, and fosters long-term engagement and project growth \cite{zhou2012make, iaffaldano2019developers, Zhou2010DeveloperFA, qiu2019going}. However, managing retention in OSS is challenging due to the many factors that influence why contributors stay or leave. This section reviews existing literature on how contributor retention is managed in OSS.

\subsection{The Complexities of Managing Contributor Retention}

Research on OSS contributor retention has examined both individual and project-level indicators. Individual metrics include commit frequency, comments, and pull request activity \cite{qiu2019going, eluri2021predicting, bao2019large}, while project factors encompass popularity and collaboration environment characteristics \cite{yamashita2016magnet, iaffaldano2019developers, zhou2012make, zhou2011does, eluri2021predicting, bao2019large}.

Existing studies indicate that contributors’ demographic attributes can influence their likelihood of disengaging, particularly when certain groups experience exclusion within project communities. For instance, \citet{trinkenreich2022women} found that women contributors are underrepresented in the OSS community and face a heightened risk of disengagement. \citet{zhang2019companies} found that affiliated developers contribute more than volunteers in the OpenStack Foundation, but this can lead to a Pareto-like phenomenon affecting an OSS project's sustainability if dominant companies withdraw. Beyond gender, other demographic factors—such as geographic region and English proficiency—may also affect retention \cite{feng2023state1, feng2023state}. While these insights highlight critical risks to retention, translating them into actionable strategies remains a persistent challenge for practitioners.

Moreover, contributor retention is shaped by a wide range of factors, requiring community managers to actively track and interpret various signals. However, they then face challenges in leveraging retention data due to its fragmented and complex nature \cite{guizani2022attracting, gray2022disengage}.
For example, one of the most popular OSS project health initiatives---the CHAOSS project \cite{CHAOSS}---defines over 50 metrics, organized into 10 categories such as Organization, Software, Contributions, Contributors, and Governance and Leadership, to comprehensively assess project health. In most cases, community managers lack an understanding of where to focus their efforts, what to prioritize, and how to act upon these metrics to manage retention challenges. How these complexities can collectively affect community managers’ day-to-day efforts remains unclear. Today, little is known about the hurdles community managers encounter when interpreting and acting upon these different data sources, leaving a gap in our understanding of how to support them effectively.

\subsection{Limitations of Existing Interventions}

Previous interventions for managing contributor retention span academic and industry solutions. Academic approaches include prototypes for engagement that indirectly support retention by helping with newcomer attraction and recognition \cite{guizani2022attracting}, community health monitoring dashboards that visualize project activities (e.g., pull requests and issue tracking) \cite{qiu2023climate}, and activity tracking systems for incubator projects \cite{ramchandran2022exploring}. Industry tools like DevStats \cite{cncfDevStats} and the Amazon OSS Dashboard \cite{amzn_oss_dashboard} offer similar monitoring capabilities. However, these solutions primarily provide retrospective insights, potentially delaying critical interventions. This reactive approach limits visibility into where proactive support is most needed for community managers.

In addition to analysis of historical data, OSS researchers have drawn inspiration from medicine and human resources disciplines, adopting prediction models such as the Cox regression to predict contributors at higher risk of disengagement based on their contribution history \cite{bao2019large, eluri2021predicting, giovanini2021leveraging}. Similarly, Kaplan-Meier analysis has been used to evaluate the survival rate of demographic groups within OSS communities \cite{qiu2019going, feng2023state}. While prediction models like Cox regression and Kaplan-Meier analysis have shown statistical promise, they remain isolated in research contexts and have not been integrated into tools or assessed for their effectiveness in helping community managers' decision making.

\subsection{Our contributions}

To bridge these gaps, our study employs a mixed-methods approach guided by the DSR paradigm to create a comprehensive understanding of the complexities of managing contributor retention in OSS. We identified community managers' challenges by triangulating findings through interviews, multi-vocal literature reviews, and surveys. Building on these insights, we curated a set of strategies designed to help OSS community managers identify and address disengagement risks, making informed decisions.  These strategies were embedded in a web-based prototype, refined through a user-centric design process involving over 100 OSS practitioners. Finally, we conducted in situ evaluations with two OSS communities to assess the practicality of these strategies in supporting retention management in OSS.

%% file: section/sec3_overall_method.tex
\section{Method Overview}

This section provides an overview of our methodology (Figure \ref{fig:method}), outlining how we answered each research question by following the DSR paradigm \cite{runeson2020design, engstrom2020software} through a mixed-methods \cite{easterbrook2008selecting, storey2024guiding}, two-phase study that addresses the problem of contributor retention in OSS.

Phase 1 - \textit{Problem definition}: We conceptualized the problem to identify the challenges in managing retention in OSS by triangulating insights from semi-structured interviews, a multi-vocal literature review \cite{garousi2019guidelines}, and an expert survey.

Phase 2 - \textit{Solution design and validation}: We designed a solution by first synthesizing strategy candidates from prior work and community input, then applying participatory, user-centric design methods to iteratively refine a prototype through multiple focus group sessions with over 100 OSS contributors at major community events. We validated the strategies in context by conducting an in situ user evaluation across two OSS communities (Pyomo and DeepSpeed).  

These two phases were guided by our two research questions as follows:

\begin{figure}[!bt]
    \centering
    \includegraphics[width = \textwidth]{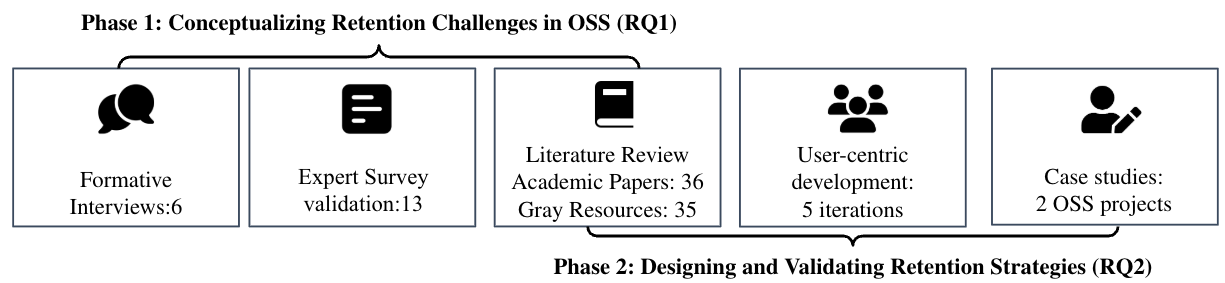}
    \caption{Method Overview}
    \label{fig:method}
%\vspace{-5mm}
\end{figure}

\textbf{(RQ1)} For problem identification, we focused on empirically surfacing and understanding the challenges related to retention management in OSS. To ensure methodological rigor, we triangulated multiple sources of evidence:  We initiated our investigation by conducting \textit{semi-structured interviews} (\faComments) with OSS community managers. These initial findings were then triangulated \cite{flick2004triangulation} through (1) \textit{a multi-vocal literature review} (\faBook) \cite{garousi2019guidelines} (academic: \cite{steinmacher2016overcoming,zhang2022turnover,lin2017developer,
lee2018one,robinson2022two,qiu2019going,zhou2016inflow,ait2022empirical,
steinmacher2015increasing,bao2019large,steinmacher2018let,
avelino2019abandonment,low2015software,schilling2012will,carillo2014s,
li2021you,trinkenreich2021please,schilling2014we,steinmacher2013newcomers,
barcomb2019episodic,yu2012empirical,vasilescu2015gender,scott2020productivity,
miller2019people,zhou2012make,eluri2021predicting,valiev2018ecosystem,
iqbal2014understanding,iaffaldano2019developers,yamashita2016magnet,
zhou2014will,tsay2014influence,midha2007retention,raman2020stress,
nassif2017revisiting,sharma2012examining} and industry
gray literature: \cite{G1,G2,G3,G4,G5,G6,G7,G8,G9,G10,%
G11,G12,G13,G14,G15,G16,G17,G18,G19,G20,%
G21,G22,G23,G24,G25,G26,G27,G28,G29,G30,%
G31,G32,G33,G34,G35})  and (2) \textit{an expert validation survey (\faPollH)}. The methodology details are explained in Section \ref{sec:challenge}.

\textbf{(RQ 2)} For solution design and validation, we continued our mixed-method approach to identify, develop, and refine strategies that support OSS community managers in diagnosing contributor engagement patterns and responding to retention challenges. First, we revisited the \textit{literature} (\faBook) used for RQ1 to compile a list of potential strategies. Then, we developed a web-based prototype that operationalized these strategies. Prototyping enabled us to move beyond theoretical discussions by giving OSS community managers an interface to interact with and assess the strategies within the context of their project. This approach enabled us to gather richer, more actionable feedback, revealing how these strategies can be integrated into existing workflows and identify potential limitations \cite{besanccon2016hybrid}. See the list of strategies in Section \ref{sec:stratagies}.

To ensure a community-driven prototype design, and in line with DSR’s emphasis on iterative development, we followed a \textit{user-centric design} (\faUsers) approach \cite{ferrario2014software} (see Figure~\ref{fig:FGD}), holding multiple rounds of focus group discussions with over 100 OSS contributors throughout the design process, detailed below:

\textsc{1. Initial Feedback from Focus Group Discussion I.} We adhered to design principles and guidelines from \citet{few2006information} to create an early prototype version. We then gathered initial feedback from community managers at an industry event (FOSSY 2023), using their insights to refine our design.
\textsc{2. Initial Implementation and Consultations with Kubernetes and Flutter.} Based on input from FOSSY 2023 participants, we consulted with OSS managers from two large-scale projects, Kubernetes and Flutter, to enhance and tailor the prototype’s features.
\textsc{3. Iterative Refinement from Focus Group Discussion II.} We organized a community session at an industry event (Linux Plumbers 2023), engaging with community managers and maintainers. Feedback from this session helped us iterate on the prototype to better align with community needs. \textsc{4. Final Feedback from Focus Group Discussion III.} In a community session at another event (OPENUK 2023), we received no further design feedback, indicating that our design was sufficiently robust to meet community expectations. The details of how we operationalize strategies in the prototype are discussed in Section \ref{sec:stratagies-ii}.

To validate the strategies in context, as emphasized in the DSR paradigm, we conducted an \textit{in situ user evaluation} (\faUserEdit) by introducing the prototype to two OSS communities with different governance structures and cultures (Pyomo \cite{pyomo2024} and DeepSpeed \cite{deepspeed2024}). We populated the prototype with their project data and invited their governance teams to evaluate whether the prototype influenced their decision-making around contributor retention. This step provided us with real-world feedback on the usefulness of the provided strategies and insights. The details of study design are explained in Section \ref{sec:evaluation}.

This study was approved by the Oregon State University Institutional Review Board (IRB) under the project titled \textit{Sustainable OSS: A Framework for Proactive Identification and Reduction of Contributor Disengagement}. All research activities involving human participants were carried out in accordance with IRB requirements, including informed consent procedures.

%% file: section/sec4_challenge.tex
\section{Challenges in Managing Contributors' Retention in OSS}
\label{sec:challenge}

\subsection{Research Methodology}

Following the DSR paradigm, we began by conceptualizing the problem space. To understand how to help community managers with contributor retention, we first investigated the challenges OSS community managers face. We employed a multiple-triangulation validation strategy \cite{flick2004triangulation}, as shown in Figure~\ref{fig:method}, starting with interviews (\faComments) to directly understand challenges from community managers. We then triangulated our findings through a multi-vocal literature review (\faBook) \cite{garousi2019guidelines}, which included gray literature and academic research, and an expert validation survey with OSS project managers (N=13) with diverse community management experiences (\faPollH).

\subsubsection{Semi-structured Interviews \faComments:}~\\

\vspace{-5mm}
\textbf{Interview Recruitment:} Guided by principles from qualitative methodology \cite{braun2021thematic, guest2006many}, we recruited participants with diverse OSS experience levels, community management tenure, and project sizes—including managers from small, medium, and large OSS communities. Table~\ref{tab:participant_demo} summarizes the participants’ demographics.
We first recruited two community managers from the authors' contacts: one from Cloud Native Computing Foundation (CNCF) and the other from Community Health Analytics in Open Source Software (CHAOSS). We then employed a snowball sampling method by asking participants to help us recruit additional participants. 
Recruitment ended after six interviews, following the principle of thematic saturation \cite{sebele2020saturation}, as no new categories of insight emerged. Starting with the fifth participant, interviews ceased to generate additional perspectives on the challenges of managing contributor retention, and the sixth interview served as a validation check.

\input{table/interview_demographics}

Among the six interviews, four were conducted on-site, and two were conducted remotely through a video conference platform \footnote{\href{https://zoom.us}{Zoom}}. The interviews lasted from 45 minutes to 1 hour, were recorded with the participants' consent (following the university-approved IRB protocol), and were then transcribed. Participants were offered a \$50 gift card as a token of appreciation for their time.

\textbf{Interviews Design:} Each session began with a brief introduction to the study. After confirming the participants' consent, we asked them to reflect on the composition and roles within their community (e.g., ``How many newcomers are there in your project?''). We used this icebreaker question to help participants recall their community's composition and contextualize the interview regarding their management experiences.

We then asked about each participant's role and responsibilities in the project. Next, we invited them to share their thoughts on contributor retention and any observations they had about contributors' disengagement. We followed up to determine whether they actively monitored or assessed retention using specific tools or metrics, taking into account their challenges. Throughout the interview, open-ended follow-up questions were actively used to encourage deeper insights into how they manage contributor retention.

\textbf{Pilot Interviews:} Before conducting interviews, we held three rounds of pilot interviews with graduate research assistants who were knowledgeable about OSS communities and had relevant experience in OSS research. Feedback from these pilot sessions helped us refine the interview questions. For example, participants suggested adding ice-breaking questions to contextualize the interview and follow-up prompts for situations in which respondents provided short answers about their experiences in managing contributor retention. They also recommended asking participants to share specific examples of times they observed or anticipated a contributor leaving the project. (Refer to the supplementary material for detailed interview scripts \cite{suppl}).

\textbf{Interview Data Analysis:} We did not separate the interviews and their analysis. Instead, after each interview, we performed an inductive thematic analysis to identify emerging themes \cite{glaser2016open, braun2006using}. We made multiple passes over the transcribed data, avoiding presupposed codes and allowing codes to emerge naturally from the content. Throughout the analysis, three authors discussed new codes, developed a preliminary codebook, and continuously evaluated the rationale for assigning specific codes. We conducted the entire procedure through continuous comparison during coding sessions and used negotiated agreements \cite{garrison2006revisiting, forman2007qualitative}.

Our analysis resulted in the identification of codes related to (1) the challenges of monitoring contributor retention in OSS, (2) methods for monitoring contributor retention, and (3) experiences of managing retention in OSS.  The three researchers agreed on 10 challenges in monitoring contributor retention (see Table~\ref{tab:monitor_challenges}).

\subsubsection{Triangulation I: Literature Review \faBook} We conducted a multi-vocal literature review to triangulate the challenges in managing contributor retention that we learned from the interviews and to investigate whether we overlooked any additional challenges.

\textbf{Scope of Literature Review:} Our literature review was designed to cover both academic and gray literature (e.g., OSS blogs), as studies have mentioned that disconnection and miscommunication between research and industry remain prevalent in software engineering communities \cite{rico2020exploring}. OSS practitioners often share their experiences on blog platforms, which academic studies might overlook. By including both literature resources, we aimed to provide a more comprehensive understanding of the challenges in managing retention in OSS.

\textbf{Keywords:} We determined a list of search keywords to ensure the relevance of the captured studies about retention in OSS. We used three studies on retention management in OSS that we were already familiar with to decide on keywords \cite{qiu2019going, zhou2016inflow, schilling2012will}. Proposed search keywords included, ``retention,'' ``disengagement,'' ``turnover,'' ``inactive'',  ``abandonment'', ``survive,'' and ``survival''. We added ``Open Source" or ``OSS" to each string to focus on the OSS community.

\textbf{Academic literature review:} We required the search keywords to appear at least once in the title, abstract, or keywords. This approach is consistent with existing systematic literature reviews in OSS research \cite{feng2024guiding}. The digital libraries (IEEE Xplore, ACM Digital Library) we used have been frequently utilized in other systematic literature reviews in software engineering \cite{ambreen2018empirical, feng2024guiding, trinkenreich2022women}. Additionally, following the recommendations of \citet{kitchenham2013systematic} for secondary studies in computer science and software engineering, we employed backward snowballing to identify any relevant studies we might have missed.

\textbf{Pilot search queries in academic digital libraries:} Before applying the search string across the digital libraries, we conducted a pilot search to validate the query. We used three control studies \cite{qiu2019going, zhou2016inflow, schilling2012will} to test the search query, verifying that these studies were retrieved.

\textbf{Academic literature list:} Our initial search after removing duplicates resulted in 256 publications. Then, the first and second authors read the titles and abstracts and only selected papers focused on retention in OSS; we only kept papers with more than five pages; those with fewer than five pages are usually not considered full research papers in top-tier venues (e.g., \cite{icse_2025, fse_2025, feng2024guiding}). When we finished filtering out papers, 24 remained.

To collect additional studies, we performed a single iteration of backward snowballing \cite{kitchenham2013systematic, wohlin2014guidelines}, looking for further studies published in journals and other conferences. This search resulted in us adding 11 papers to our literature list, resulting in a final total of 35 academic papers.

To assess the completeness of our list, we performed a completeness check \cite{feng2024guiding} on the three most recent studies from our list: \cite{calefato2022will,trinkenreich2023belong,qiu2023climate}. This step ensured we did not miss any relevant studies, as we only performed a single iteration of backward snowballing, and keyword search results may have missed some related studies. The first two authors independently reviewed every reference in these three papers, listing those relevant to contributor retention. The two authors then reviewed each other's lists and reached a negotiated agreement. In the end, we found 35 academic studies, with only one missing from our list \cite{raman2020stress}. Thus, our completeness ratio is 99\%, with one additional paper being added to our original list, resulting in a final total of 36 academic papers \cite{steinmacher2016overcoming,zhang2022turnover,lin2017developer,
lee2018one,robinson2022two,qiu2019going,zhou2016inflow,ait2022empirical,
steinmacher2015increasing,bao2019large,steinmacher2018let,
avelino2019abandonment,low2015software,schilling2012will,carillo2014s,
li2021you,trinkenreich2021please,schilling2014we,steinmacher2013newcomers,
barcomb2019episodic,yu2012empirical,vasilescu2015gender,scott2020productivity,
miller2019people,zhou2012make,eluri2021predicting,valiev2018ecosystem,
iqbal2014understanding,iaffaldano2019developers,yamashita2016magnet,
zhou2014will,tsay2014influence,midha2007retention,raman2020stress,
nassif2017revisiting,sharma2012examining} .

\textbf{Gray literature review:} We searched the top 50 OSS blog platforms from the list of ``Best Open Source Blogs and Websites'' on Feedly.com \cite{feedly2024}. These blog platforms included Open Source Initiative \cite{opensourceblog2024}, Google Open Source Blog \cite{googleopensourceblog2024}, Linux.com \cite{linuxcom2024}, the Software Freedom Conservancy blog \cite{sfconservancyblog2024}, Fedora People \cite{fedorapeople2024}, and other foundation-based platforms. We removed 18 platforms from the list as they were either project-specific, solely released project updates, or personal web pages (See the Supplementary Materials for the list of gray literature platforms \cite{suppl}). Unlike larger OSS blog platforms, which typically undergo some review and reflect common or widely accepted practices within the community, personal web pages can present unverified opinions, making them less reliable. We used the exact search string that we used to find academic literature, which resulted in 35 blog articles \cite{G1,G2,G3,G4,G5,G6,G7,G8,G9,G10,%
G11,G12,G13,G14,G15,G16,G17,G18,G19,G20,%
G21,G22,G23,G24,G25,G26,G27,G28,G29,G30,%
G31,G32,G33,G34,G35}.

The first two authors independently reviewed papers and blog content to find any challenges related to monitoring contributors' retention. We held weekly meetings to present our findings and discuss the identified challenges until we reached an agreement. The challenges agreed upon in this open coding iteration were added to the interview results.

\subsubsection{Triangulation II: Expert Validation survey \faPollH}  We then conducted an expert evaluation survey to triangulate the challenges of managing retention.

\textbf{Survey design:} Our survey contained 14 questions, including closed- and open-ended questions. The survey started with three demographic questions regarding gender, OSS experience, and experience as OSS managers. The following 10 questions aimed to understand the participants' agreement with the challenges we identified in managing contributors' retention. The final question was open-ended, asking participants to list any challenges not listed in the survey.

\textbf{Pilot survey:} We conducted five pilot studies with graduate students (n=3) and OSS contributors (n=2). After each pilot study, we collected feedback, which we used to refine the survey. For example, pilot study participants suggested adding explanations and short examples for each challenge. See the supplementary materials for the survey questions \cite{suppl}.

\textbf{Participant recruitment:} Once we finalized the survey, we began the recruitment process by contacting five community managers new to the study (i.e., didn't participate in the earlier interviews) and had more than five years of experience managing OSS communities. These community managers helped us recruit 15 participants. After removing seven incomplete answers, we had 13 valid responses. The demographics of the survey participants are shown in  Table \ref{tab:survey_responses}. Nine out of 13 participants have experience managing large projects (more than 100 contributors), and five mentioned having managed small to large communities. The diverse OSS and management experiences provided a foundation for effectively validating the identified challenges.

\input{table/survey_response}

\textbf{Survey analysis:} We quantitatively analyzed the survey's closed-ended questions to understand participants' level of agreement on the identified challenges in contributor retention monitoring. We did not receive any responses from the open-ended questions regarding new challenges.

In summary, our analysis of the interviews revealed 10 challenges. During the literature review, we found no new challenges; 10 out of 10 were validated from research studies, and nine (out of 10) challenges were validated from blog posts, as shown in Table \ref{tab:monitor_challenges}. The survey responses helped us triangulate the 10 challenges, as shown in the last column of Table \ref{tab:monitor_challenges}. Dark green indicates ``strongly agree'', and light green indicates ``agree'', with each challenge receiving support from more than half of the participants.

\subsection{Results}

Here, we unpack the challenges of managing contributor retention, which we group into four categories: (1) workload and time constraints, (2) fragmented contributor and contribution tracking, (3) contributor disengagement, and (4) data privacy. Table \ref{tab:monitor_challenges} presents these challenges. For each challenge, the table includes its definition, an example, and supporting evidence from our triangulation validation.

\input{table/monitor_challenges}

\subsubsection{Workload and Time constraints}

\textbf{C1. Lack of a Retention Management Mechanism:} Participants described the absence of a mechanism or infrastructure to support contributor retention efforts. Instead of relying on streamlined or automated tools to infer disengagement, community managers have to resort to manual, piecemeal methods: scanning spreadsheets [P2], reviewing mailing lists [P1, P6], or relying on personal familiarity with contributors [P2, P4]. As one participant noted, \textit{“I derive a lot of the data manually”} [P3]. Relying on manual processes is impractical for large-scale communities with thousands of contributors \cite{avelino2019abandonment, zhou2012make, lin2017developer}. As OSS communities grow, managers struggle to track and combine diverse data sources and metrics to gauge engagement and retention accurately \cite{steinmacher2013newcomers, ait2022empirical, steinmacher2015social}.

\textbf{C2. Overwhelming Responsibilities for OSS Community managers} have been widely reported in OSS communities \cite{G2, G3, G4, G5, G6, G7, G8, G9, G10, G11, G12, G13, G14, G31}. One Blog noted, \textit{``Lack of time was cited as the leading reason ‘not to contribute' and a motivation to leave a community''} \cite{G6}. Similarly, one interview participant mentioned they were \textit{``already probably putting in somewhere between a 50–70 hour work week''} [P3], while another noted that \textit{``doing the interpersonal work communities need to grow is overwhelming''} [P2].

\textbf{C3. Challenges of Managing Turnover:} Turnover within OSS communities can undermine sustainability by causing knowledge loss and increasing onboarding costs for newcomers \cite{steinmacher2016overcoming, li2021you, iqbal2014understanding, trinkenreich2021please, lin2017developer}. \textit{``87\% of hiring managers experience difficulties recruiting enough open source talent, similar to last year when 89\% reported challenges in finding the right mix of experience and skills''} \cite{G18}. Our interview participants similarly reported that managing turnover is challenging \textit{``because even if someone isn't a `maintainer', if they've been responsible for most of the code changes in our database backend driver, losing them abruptly and unexpectedly is still going to be a problem for the project''} [P3]. Consequently, it can be difficult to identify candidates with comparable expertise to replace contributors who leave the project. Such incidents may discourage community managers, which can in turn force them to leave the community \cite{barcomb2019episodic, zhang2022turnover}.

\textbf{C4. Time-consuming to Engage and Grow the Contributor Base:} P2 described the personal effort involved in keeping contributors engaged: \textit{``Usually, if someone has been active and then goes inactive, I will reach out and say, ‘Hey, what's up? No pressure, just wanted to reach out and see how you're doing.' And then, at that point, people will respond nine times out of 10, and someone will explain''} [P2]. However, such individualized interaction is not feasible in day-to-day practice, especially for large communities; actively engaging with and cultivating a vibrant contributor base is time-consuming \cite{steinmacher2018let}.  \textit{``Make sure you're prepared to spend the time doing community building and interacting with people to make it truly shine. We struggled with that early on, and it really hit us hard''} [P5].

\subsubsection{Fragmented Retention Data}

\textbf{C5. Challenges in Tracking Project Data:} Project retention indicators take many forms, such as the number of active contributors, the number of contributors who have left, the number of newcomers joining, and are often scattered across multiple channels, including pull requests and issues. Because of this, community managers find it challenging to assess the overall retention health of their projects \textit{``Sometimes I need to see the overall trend across a project or several projects at once if they're related to each other.''} [P6].  This process demands substantial effort because it involves collecting fragmented project data and aggregating them \cite{bao2019large, li2021you, yu2012empirical, raman2020stress}.

\textbf{C6. Challenges in Aggregating Contributor Data:} Delving into details about individual contributors, such as understanding their trajectory, identifying where they are stalling, and recognizing the types of work they excel at, is difficult, but important to find and support at-risk contributors. No existing tool provides such insight \textit{``There are many other times when I want to see the overall trend for a particular person''} [P6]. Such inability to unify contributor-specific data impedes efforts to diagnose disengagement and proactively support at-risk members \cite{G1, G23}.

\textbf{C7. Challenges in Acknowledging Hard-to-Track Contributions:} Appropriately recognizing various forms of contribution in OSS is difficult, mainly when contribution tracking relies on repository visible metrics (e.g., commits, pull requests) \cite{raman2020stress, bao2019large}. The concept of ``glue work'', first introduced by Tanya Reilly \cite{reilly_glue}, refers to the essential yet frequently overlooked contributions in software development that hold teams and projects together. This issue is especially prevalent in OSS communities, where the dynamic nature of volunteer-driven collaboration can obscure the importance of these crucial efforts (e.g., PR reviews or mentoring). These types of glue work often remain in the shadow of feature code development  \cite{G11, G24, G25}.

\subsubsection{Contributor disengagement}

\textbf{C8. Challenges in Fostering Inclusive Collaboration:} While fostering a welcoming environment for contributors with diverse backgrounds is beneficial \cite{qiu2019going, trinkenreich2021please, vasilescu2015gender, foundjem2021onboarding}, it places additional demands on community managers who are often already overwhelmed. Studies show that contributors may face barriers rooted in their different backgrounds, such as their English proficiency, cultural differences, and gender \cite{qiu2019going, trinkenreich2021please, vasilescu2015gender}. These concerns have drawn increasing attention across OSS communities \cite{G2, G8, G15, G21, G24, G32}.  \textit{``I don't know the demographics. It is highly international; there are people from all over'' } [P6].

\textbf{C9. Challenges in Anticipating Contributor Attrition:}  Anticipating contributor attrition involves more than observing overarching trends; it requires analyzing each contributors' history of participation \cite{zhang2022turnover, lin2017developer, lee2018one, robinson2022two, qiu2019going, zhou2016inflow, bao2019large, avelino2019abandonment}. Prior research suggests that examining a contributor's historical activities can offer early indicators of potential departure \cite{zhou2016inflow, schilling2012will}. However, systematically monitoring this information is a challenging task for practitioners. \textit{``You can analyze the mailing list and compare it with the commit logs and the bug tracker to see who has been active for a long time, but I've observed many of the same names remaining active over the years''} [P6]. In some large projects, managers only realize a contributor has left after noticing the project encountered a failure \cite{li2021you}.

\subsubsection{Data Privacy}

\textbf{C10. Challenges in Ensuring Data Privacy While Tracking:} When community managers dedicate considerable effort to aggregating contributor data for retention tracking, data privacy becomes critical. Understanding contributor backgrounds is essential for fostering inclusivity and improving retention strategies. However, this process introduces privacy challenges, especially when data inaccuracies exacerbate existing issues \cite{qiu2019going, lin2017developer}.  \textit{``But the problem you encounter is that it delves into personal data about people, right? So, it becomes really hard when you start measuring this and considering who gets access to it and how you ensure people's private information remains private. Therefore, in some cases, we don't measure it because we're too concerned about doing it''} [P1].

\subsubsection{Triangulation}
To validate the interview results, we carried out two forms of data triangulation: (1) a literature review of academic and gray literature and (2) a survey.

\textit{Triangulation with Literature Review:} All 10 identified challenges were also discussed in academic literature, with the most frequently discussed challenges being C1 (lack of retention mechanism), C2 (overwhelming responsibilities), and C9 (anticipating contributor attrition). The gray literature supported 9 out of the 10 challenges; C10 (ensuring data privacy while tracking) was not supported in our gray literature review. The top-3 reported challenges in the gray literature were C2, C3, and C9. This shows that both academic and gray literature are well aware of the problem of maintainer burnout because of being overwhelmed with responsibilities (C2) as well as challenges in being able to predict contributor attrition (C9).

\textit{Triangulation with Survey Responses:} For all 10 challenges, more than 50\% of survey respondents agreed  (strong agreement/agreement shown by dark/light green bars) that these challenges exist in their projects. The top three highest rates of agreement were for challenges C3 (managing turnover), C4 (engaging and growing the contributor base), and C8 (fostering inclusive collaboration). The list of the top three is different from what we found in the academic literature, but this is likely because the challenges are project-dependent.

%% file: table/interview_demographics.tex
\begin{table}[!htbp]
\caption{Demographic information of interview participants.}
\resizebox{5in}{!}{
\begin{tabular}{lllll}
\rowcolor[HTML]{EFEFEF} 
\textbf{ID}              & \textbf{Gender}                & \textbf{OSS Experience}              & \textbf{Management Experience}              & \textbf{Current Project Size*}              \\
P1                       & Woman                          & More than 10 years                   & Between 3 and 5 years                       & Medium                                      \\
\rowcolor[HTML]{EFEFEF} 
P2                       & Woman                          & More than 10 years                   & Between 5 and 10 years                      & Medium                                      \\
P3                       & Man                            & Between 3 and 5 years                & Less than 1 year                            & Large                                       \\
\rowcolor[HTML]{EFEFEF} 
P4                       & Perfer not to say              & Between 3 and 5 years                & Between 1 and 3 years                       & Small                                       \\
P5                       & Man                            & More than 10 years                   & More than 10 years                          & Small                                       \\
\rowcolor[HTML]{EFEFEF} 
P6                       & Man                            & More than 10 years                   & Between 5 and 10 years                      & Large                                       \\
\multicolumn{5}{l}{\begin{tabular}[c]{@{}l@{}}Project Size: Small projects (1-25 contributors), Medium projects (26 - 100 contributors), \\ Large projects (>100 contributors)\end{tabular}}
\end{tabular}}
\label{tab:participant_demo}
%\vspace{-3mm}
\end{table}

%% file: table/survey_response.tex
\begin{table}[!tbp]
\caption{Expert survey participant demographics}
\resizebox{3.5in}{!}{
\begin{tabular}{llll}
\rowcolor[HTML]{EFEFEF} 
\textbf{ID}                                 & \textbf{Gender}                                  & \textbf{OSS experience}                                       & \textbf{Managing experience}                                \\
SP1                                         & Man                                              & More than 10 years                                            & SML                                                         \\
\rowcolor[HTML]{EFEFEF} 
SP2                                         & Woman                                            & 3 years but less than 5 years                                 & S                                                           \\
SP3                                         & Woman                                            & 3 years but less than 5 years                                 & SM                                                          \\
\rowcolor[HTML]{EFEFEF} 
SP4                                         & Man                                              & More than 10 years                                            & L                                                           \\
SP5                                         & Man                                              & More than 10 years                                            & SML                                                         \\
\rowcolor[HTML]{EFEFEF} 
SP6                                         & Man                                              & 5 years but less than 10 years                                & SML                                                         \\
SP7                                         & Man                                              & More than 10 years                                            & SML                                                         \\
\rowcolor[HTML]{EFEFEF} 
SP8                                         & Man                                              & More than 10 years                                            & S                                                           \\
SP9                                         & Man                                              & More than 10 years                                            & L                                                           \\
\rowcolor[HTML]{EFEFEF} 
SP10                                        & Woman                                            & More than 10 years                                            & SML                                                         \\
SP11                                        & Man                                              & 1 year but less than 3 years                                  & M                                                           \\
\rowcolor[HTML]{EFEFEF} 
SP12                                        & Prefer not to say                                & 5 years but less than 10 years                                & L                                                           \\
SP13                                        & Man                                              & 5 years but less than 10 years                                & L                                                           \\
\rowcolor[HTML]{EFEFEF} 
\multicolumn{4}{l}{\cellcolor[HTML]{EFEFEF}\begin{tabular}[c]{@{}l@{}}Managing experience: Small projects (1-25 contributors),\\ Medium projects (26 - 100   contributors), Large projects (>100 contributors)\end{tabular}}
\end{tabular}}
\label{tab:survey_responses}
\end{table}

%% file: table/monitor_challenges.tex
\begin{table}[!tbp]
\caption{Challenges of managing contributor retention.}
\resizebox{\textwidth}{!}{
\begin{tabular}{lllllll}
\rowcolor[HTML]{EFEFEF} 
\textbf{ID} & \textbf{\begin{tabular}[c]{@{}l@{}}Challenges in \\ managing retention\end{tabular}}                     & \textbf{\begin{tabular}[c]{@{}l@{}}Example (It is \\ challenging to...)\end{tabular}}                                                                              & \textbf{\begin{tabular}[c]{@{}l@{}}Participant \\ ID\end{tabular}} & \multicolumn{1}{c}{\cellcolor[HTML]{EFEFEF}\textbf{Academic literature}}                                                                                                                                                                                                                                                & \multicolumn{1}{c}{\cellcolor[HTML]{EFEFEF}\textbf{Gray Literature}}                                                                                                                                                                                                                                                                                                & \multicolumn{1}{c}{\cellcolor[HTML]{EFEFEF}\textbf{Survey (N=13)*}} \\
\multicolumn{7}{c}{Workload and time Constraints}                                                                                                                                                                                                                                                                                                                                                                                                                                                                                                                                                                                                                                                                                                                                                                                                                                                                                                                                                                                                                                                                                      \\
\rowcolor[HTML]{EFEFEF} 
C1          & \begin{tabular}[c]{@{}l@{}}Lack of a Retention \\ Management \\ Mechanism\end{tabular}                   & \begin{tabular}[c]{@{}l@{}}manage retention \\due to lack of tools \\ for diagnosis and \\insight.\end{tabular} & \begin{tabular}[c]{@{}l@{}}P1, P2, P3, \\ P4, P5, P6\end{tabular}  & \begin{tabular}[c]{@{}l@{}}\cite{lin2017developer, ait2022empirical, steinmacher2018let}\\ \cite{avelino2019abandonment, li2021you, steinmacher2013newcomers}\\ \cite{zhou2012make, lin2017developer}\end{tabular}                                                                                                      & \cite{G29}                                                                                                                                                                                                                                                                                                                              & \fivepointlikert{31}{23}{15}{8}{23}                                 \\
C2          & \begin{tabular}[c]{@{}l@{}}Overwhelming \\ Responsibilities\end{tabular}                                 & \begin{tabular}[c]{@{}l@{}}allocate time effectively \\ to balance responsibilities \\ while monitoring retention.\end{tabular}                                    & \begin{tabular}[c]{@{}l@{}}P2, P3, P4,\\ P5, P6\end{tabular}       & \begin{tabular}[c]{@{}l@{}}\cite{steinmacher2016overcoming, li2021you, trinkenreich2021please}\\ \cite{iqbal2014understanding, lin2017developer, barcomb2019episodic}\\ \cite{schilling2012will, avelino2019abandonment, nassif2017revisiting}\\ \cite{bao2019large, lin2017developer, zhang2022turnover}\end{tabular}  & \begin{tabular}[c]{@{}l@{}}\cite{G2, G3, G4}\\ \cite{G5, G6, G7}\\ \cite{G8, G9, G10}\\ \cite{G11, G12, G13}\\ \cite{G14, G31, G33}\end{tabular} & \fivepointlikert{69}{8}{15}{0}{8}                                   \\
\rowcolor[HTML]{EFEFEF} 
C3          & \begin{tabular}[c]{@{}l@{}}Challenges of \\ Managing \\ Turnover\end{tabular}                            & \begin{tabular}[c]{@{}l@{}}identify candidates with \\ equivalent skills to mitigate \\ the impact of turnover.\end{tabular}                                       & \begin{tabular}[c]{@{}l@{}}P2, P3, P4,\\ P5, P6\end{tabular}       & \cite{steinmacher2018let}                                                                                                                                                                                                                                                                                               & \begin{tabular}[c]{@{}l@{}}\cite{G15, G16, G17}\\ \cite{G18, G19, G20} \\ \cite{G26, G30, G32}\end{tabular}                                                                                                                                     & \fivepointlikert{54}{38}{8}{0}{0}                                   \\
C4          & \begin{tabular}[c]{@{}l@{}}Time-consuming to \\ Engage and Grow \\ the Contributor Base\end{tabular}     & \begin{tabular}[c]{@{}l@{}}dedicate sufficient time \\ to interact with and support \\ contributors effectively.\end{tabular}                                      & \begin{tabular}[c]{@{}l@{}}P1, P2, P3, \\ P5, P6\end{tabular}      & \cite{steinmacher2018let}                                                                                                                                                                                                                                                                                               & \cite{G35}                                                                                                                                                                                                                                                                                                       & \fivepointlikert{54}{38}{0}{8}{0}                                   \\
\rowcolor[HTML]{EFEFEF} 
\multicolumn{7}{c}{\cellcolor[HTML]{EFEFEF}\textbf{Fragmented Contributor and Contribution Tracking}}                                                                                                                                                                                                                                                                                                                                                                                                                                                                                                                                                                                                                                                                                                                                                                                                                                                                                                                                                                                                                                  \\
C5          & \begin{tabular}[c]{@{}l@{}}Challenges in \\ Tracking \\ Project Data\end{tabular}                        & \begin{tabular}[c]{@{}l@{}}gain a clear view of \\ overall retention trends \\ within the project.\end{tabular}                                                    & P3, P6                                                             & \begin{tabular}[c]{@{}l@{}}\cite{bao2019large, li2021you, yu2012empirical}\\ \cite{raman2020stress}\end{tabular}                                                                                                                                                                                                        & \cite{G22, G23, G28}                                                                                                                                                                                                                                                                                     & \fivepointlikert{39}{15}{31}{8}{8}                                  \\
\rowcolor[HTML]{EFEFEF} 
C6          & \begin{tabular}[c]{@{}l@{}}Challenges in \\ Tracking \\ Contributor Data\end{tabular}                    & \begin{tabular}[c]{@{}l@{}}aggregate and analyze \\ contributor activities \\ and engagement to identify \\ at-risk members.\end{tabular}                                               & P2, P3, P6                                                         & \begin{tabular}[c]{@{}l@{}}\cite{bao2019large, li2021you, yu2012empirical}\\ \cite{raman2020stress}\end{tabular}                                                                                                                                                                                                        & \begin{tabular}[c]{@{}l@{}}\cite{G1, G23, G27}\\ \cite{G29}\end{tabular}                                                                                                                                                                                                                                                                         & \fivepointlikert{31}{23}{15}{8}{23}                                 \\
C7          & \begin{tabular}[c]{@{}l@{}}Challenges in \\ Acknowledging \\ Hard-to-Track \\ Contributions\end{tabular} & \begin{tabular}[c]{@{}l@{}}identify and recognize \\ untracked contributions, \\ such as marketing efforts.\end{tabular}                                           & \begin{tabular}[c]{@{}l@{}}P1, P2, P3,\\ P4\end{tabular}           & \cite{raman2020stress, bao2019large}                                                                                                                                                                                                                                                                                    & \cite{G11, G24, G25}                                                                                                                                                                                                                                                                        & \fivepointlikert{46}{23}{31}{0}{0}                                  \\
\rowcolor[HTML]{EFEFEF} 
\multicolumn{7}{c}{\cellcolor[HTML]{EFEFEF}\textbf{Managing Diversity and Contributor Retention Risks}}                                                                                                                                                                                                                                                                                                                                                                                                                                                                                                                                                                                                                                                                                                                                                                                                                                                                                                                                                                                                                                \\
C8          & \begin{tabular}[c]{@{}l@{}}Challenges in \\ Fostering Inclusive \\ Collaboration\end{tabular}            & \begin{tabular}[c]{@{}l@{}}evaluate the demographics \\ of contributors at risk of \\ disengagement.\end{tabular}                                                  & \begin{tabular}[c]{@{}l@{}}P1, P2, P3, \\ P4, P5, P6\end{tabular}  & \cite{qiu2019going, trinkenreich2021please, vasilescu2015gender}                                                                                                                                                                                                                                                        & \begin{tabular}[c]{@{}l@{}}\cite{G10, G17, G21}\\ \cite{G24, G26, G27}\\ \cite{G32, G33, G34}\end{tabular}                                                                                                         & \fivepointlikert{39}{46}{15}{0}{0}                                  \\
\rowcolor[HTML]{EFEFEF} 
C9          & \begin{tabular}[c]{@{}l@{}}Challenges in \\ Anticipating \\ Contributor \\ Attrition\end{tabular}        & \begin{tabular}[c]{@{}l@{}}track contributor \\ activity to identify \\ early signs of \\ disengagement.\end{tabular}                                              & P2, P3, P6                                                         & \begin{tabular}[c]{@{}l@{}}\cite{zhang2022turnover, lin2017developer,lee2018one}\\ \cite{robinson2022two, qiu2019going, zhou2016inflow}\\ \cite{bao2019large, avelino2019abandonment, schilling2012will}\\ \cite{barcomb2019episodic, yu2012empirical, scott2020productivity}\\ \cite{eluri2021predicting}\end{tabular} & \begin{tabular}[c]{@{}l@{}}\cite{G15, G16, G30}\\ \cite{G32, G17, G18} \\ \cite{G19, G20}\end{tabular}                                                                                                                                  & \fivepointlikert{15}{39}{23}{8}{15}                                 \\
\multicolumn{7}{c}{\textbf{Data Privacy}}                                                                                                                                                                                                                                                                                                                                                                                                                                                                                                                                                                                                                                                                                                                                                                                                                                                                                                                                                                                                                                                                                              \\
\rowcolor[HTML]{EFEFEF} 
C10         & \begin{tabular}[c]{@{}l@{}}Challenges in \\ Ensuring Data Privacy \\ While Tracking\end{tabular}         & \begin{tabular}[c]{@{}l@{}}ensure the confidentiality \\ of contributor data during \\ tracking processes\end{tabular}                                             & P1, P3                                                             & \cite{qiu2019going, alge2006information, zhou2014will}                                                                                                                                                                                                                                                                  & -                                                                                                                                                                                                                                                                                                                                                                   & \fivepointlikert{39}{15}{46}{0}{0}                                  \\
\multicolumn{7}{l}{\begin{tabular}[c]{@{}l@{}}The dark green section represents the percentage of participants who ``Strongly Agree,'' while the light green section shows those who ``Agree.''  \\ The gray section corresponds to ``Neutral,'' the light red corresponds to ``Disagree,'' and the red section to ``Strongly Disagree.''\end{tabular}}                                                                                                                                                                                                                                                                                                                                                                                                                                                                                                                                                                                                                                                                                                                                                                               
\end{tabular}}
\label{tab:monitor_challenges}
\end{table}

%% file: section/sec5_stratagies.tex
\section{Strategies in managing contributors retention}
\label{sec:stratagies}

This section introduces strategies to support managing contributor retention, derived from a review of 36 studies on OSS retention. These strategies highlight practices and approaches that community managers can adopt to improve contributor engagement. As part of our DSR process, we continued to follow a triangulated, iterative build–evaluate cycle \cite{runeson2020design, engstrom2020software} to validate the practicality and helpfulness of the synthesized strategies. Figure~\ref{fig:method} provides an overview of our process, which comprises: (1) a literature review (\faBook: this section), (2) user-centric design approach (\faUsers: Section \ref{sec:stratagies-ii}), and (3) case studies with future users in their OSS communities (\faUserEdit: Section \ref{sec:evaluation}).  In this section, we discuss how we identified strategies from the literature review.

\subsection{Method-literature review (\faBook)}

To identify potential strategies for the challenges we have identified, we revisited the set of 36 studies related to OSS retention from Section \ref{sec:challenge}. We reused the search results from Section \ref{sec:challenge} since the search keywords used to identify these papers were not limited to only challenges but included all aspects of retention in OSS. We did not include gray literature at this stage because we plan to directly engage with OSS community managers to gather their experiences and opinions, thereby gaining broader, real-world insights into how these strategies can be practically scaled.

Each of these studies had focused on one or set of challenges in retention management and proposed recommendations for practitioners. For instance, some studies emphasize the importance of closely monitoring and engaging newcomers to enhance retention rates \cite{steinmacher2016overcoming, lee2018one, qiu2019going, zhou2016inflow, steinmacher2015increasing, steinmacher2015increasing, steinmacher2018let, avelino2019abandonment, schilling2012will, zhou2012make, eluri2021predicting, steinmacher2015increasing, eluri2021predicting, zhou2014will, gerosa2021shifting}. Other studies provide information to community managers that aid in retention management \cite{qiu2019going, steinmacher2018let, steinmacher2013newcomers, zhou2012make, valiev2018ecosystem}. Additionally, several studies have explored using statistical models to predict the likelihood of contributors leaving a project, offering community managers the necessary insights to take proactive measures \cite{zhang2022turnover, lin2017developer, robinson2022two, zhou2016inflow, ait2022empirical, avelino2019abandonment, low2015software, eluri2021predicting, valiev2018ecosystem, lin2017developer}.

We conducted another round of open coding protocol on these 36 academic studies in our primary list to focus on potential strategies for managing retention. During the analysis, each emerging code was compared to the existing codes to determine whether it represented a separate category or was a subset of an existing code. The first three authors applied this procedure via continuous comparison throughout the coding sessions \cite{garrison2006revisiting, forman2007qualitative}. Any disagreements during the sessions were resolved through a negotiated agreement to maintain reliability \cite{garrison2006revisiting}. In total, we synthesized nine strategies (Table \ref{tab:strategies}).

\subsection{Result}

\input{table/strategies}

To address the challenges identified in Section~\ref{sec:challenge}, we synthesized nine strategies that OSS community managers can use to diagnose, track, and respond to retention-related issues. Table~\ref{tab:strategies} summarizes these strategies and maps them to the corresponding challenges.  These strategies are grouped into two categories: (1) Community Engagement and Retention Analytics and (2) Project and Contributor Management.  For each strategy, the table presents a description, its function (e.g., diagnose, reduce workload, and compliance). For example, S1 (Track Project Engagement) monitors contributor activity and retention trends to diagnose potential emerging disengagement.

\subsubsection{Community Engagement and Retention Analytics} Strategies in this category focus on creating statistical analysis to help community managers take proactive action.

\textbf{S1. Track Project Engagement:} 
Understanding engagement is essential for community managers to manage retention, as it serves a diagnostic function by surfacing early signs of disengagement \cite{zhang2022turnover, lin2017developer, sharma2012examining}. For example, \citet{schilling2014we} mentioned the importance of monitoring turnover rates, which helps assess whether the project's retention rates are healthy. \citet{zhang2022turnover} recommended tracking the number of newcomers and members who left, while \citet{lin2017developer} emphasized measuring the average tenure days to understand retention. Similarly, \citet{sharma2012examining} and \citet{qiu2019going} analyzed retention using contribution activities. Together, these metrics can provide a comprehensive mechanism to track project engagement and support community managers \cite{ait2022empirical, bao2019large, schilling2014we, eluri2021predicting}. However, extracting this data can be time-consuming, \textit{``...I mean a lot of it depends on me as the community manager keeping track of people which is hard...''} [P2]. Additionally, different metrics need to be actively monitored, many of which are fragmented across multiple locations (e.g., issues, pull requests). Using this strategy therefore requires an automated monitoring system.

\textbf{S2. Track Newcomers:} A key way to manage retention is to onboard newcomers and track their progress to diagnose early signs of contributor attrition
\cite{steinmacher2016overcoming, lee2018one, zhou2016inflow, zhou2014will, eluri2021predicting, zhou2012make, schilling2012will}. 
For example, \citet{steinmacher2016overcoming} found that mentors help newcomers integrate into the project's culture, which requires them to monitor newcomers' contributions and identify areas where support may be needed \cite{steinmacher2015increasing, avelino2019abandonment}. Similarly, \citet{feng2024guiding} emphasized that effective onboarding strategies rely on understanding newcomers' specific challenges. \citet{lee2018one} mentioned that this process depends on how well the organization can track not only who the newcomers are but also their activities. This is a challenge because of the fragmented nature of OSS tools and data sources (e.g., relevant data can be located across issue trackers, code repositories, and discussion boards), which makes it difficult to piece together a cohesive picture of newcomers’ activities and progress   \cite{feng2025multifaceted, feng2024guiding}.

\textbf{S3. Track Individual Contributions:} Tracking each contributor's activity, not only newcomers, serves as a diagnostic function by providing insights into individual engagement and productivity within the community \cite{zhou2012make, eluri2021predicting, zhou2014will, zhang2022turnover, scott2020productivity, schilling2012will, carillo2014s, li2021you}. For example, \citet{zhou2012make} and \citet{qiu2019going} highlighted that individual performance and engagement histories significantly influence retention. OSS projects can build stronger relational ties by paying close attention to the activities and social dynamics of individual developers \cite{liang2022understanding, zhou2014will, schilling2012will, li2021you, carillo2014s}. However, current platforms like GitHub and other OSS collaboration tools often disaggregate this information, making it challenging for managers to track contributors effectively. Most existing tools, such as DevStats \cite{cncf_devstats}, Amazon Dashboard \cite{amzn_oss_dashboard}, and Climate Coach \cite{qiu2023climate}, rarely provide aggregated data on individual contributors.

\textbf{S4. Predict Contributor Attrition:} Tracking contributors' attrition is important to help with retention \cite{zhou2016inflow, ait2022empirical, lin2017developer, low2015software, eluri2021predicting, valiev2018ecosystem, zhang2022turnover, lin2017developer, robinson2022two}. For example, \citet{phillips2015high} mentioned that tracking attrition of high-impact members who choose to leave can help identify areas where attraction/retention needs to be proactively managed. Research studies have proposed statistical modeling approaches to analyze the risk of contributors leaving projects \cite{zhou2016inflow, qiu2019going, eluri2021predicting, yu2012empirical}. However, these approaches have yet to be integrated into practical tools for community managers.

\textbf{S5. Understand the Impact of Attrition:}
Diagnosing project health by understanding the impact of attrition helps to ensure that the project remains competitive and retains its most needed contributors \cite{yu2012empirical, lin2017developer, zhang2022turnover, yu2012empirical}. For example, studies \cite{behfar2018knowledge, rashid2017exploring} emphasize the need to identify which parts of a project are most affected by contributor disengagement, enabling managers to understand better and mitigate the effects of attrition. When contributors leave, community managers must leverage data on existing contributors' contributions to identify candidates to take over tasks, thereby minimizing the negative impact of turnover \cite{kumar2022impact, dawn2013talent}.

\textbf{S6. Promote a Welcoming Environment:} 
Studies have found that a non-welcoming environment negatively impacts retention \cite{qiu2019going, rodriguez2021perceived, offermann2013inclusive, trinkenreich2021please, vasilescu2015gender, valiev2018ecosystem}. One practical approach to creating a welcoming, collaborative environment is to understand contributors' backgrounds \cite{sparkman2019exploring, qiu2019going}, which supports early diagnosis of exclusion risks and helps create conditions for long-term engagement. For instance, \citet{trinkenreich2023belong} highlighted that when contributors’ varied backgrounds and contributions are acknowledged, it can enhance collaboration and retention in OSS projects. Similarly, being aware of contributors who are non-native English speakers can help projects recognize the need for multi-language documentation \cite{feng2023state1}. However, collecting background information (e.g., where contributors come from and which affiliations they belong to) is challenging because it is not readily available and often requires community managers to manually gather and analyze data from multiple sources \cite{qiu2019going, vasilescu2015gender, lin2017developer}.

\subsubsection{Project and Contributor Management} Strategies in this category emphasize automated management tools that enable streamlined processes for both contributors and community managers.

\textbf{S7. Ensure Privacy Practices:} While gathering contributor background information can support inclusion initiatives, privacy concerns are a challenge. Efforts to anonymize developer information can inadvertently lead to misclassifications, potentially causing distress among contributors \cite{qiu2019going, lin2017developer, thelwall2011privacy}. Moreover, if such information is unavailable, using tools to aggregate data raises questions about how to protect individual privacy \cite{thelwall2011privacy, lin2016recognizing}. For instance, \citet{wong2019human} and \citet{andrade2020privacy} reported cases of contributors' email addresses and affiliations leaking, raising significant concerns for many organizations.  Balancing the need for contributor information with privacy protections requires careful consideration \cite{zimmer2020but}.

\textbf{S8. Automated Notifications to Monitor Community Health} could help support community managers and reduce their workload \cite{steinmacher2016overcoming}. Such automated notifications can provide timely updates on problems or metrics that require attention \cite{steinmacher2016overcoming, steinmacher2018let, iqbal2014understanding}. 
For example, \citet{lin2023automatic} found that using automated notifications to highlight updates helps users quickly focus on important information, reducing the need for constant manual monitoring. However, while dashboards like DevStats \cite{cncf_devstats} offer insights on community activity, they can sometimes add to the cognitive load when community managers are already overwhelmed by routine tasks \cite{preece2009reader, kraut2012building}.

\textbf{S9. Utilizing tools for Contributor Engagement} is important to help reduce community managers' workload. Automated processes, such as onboarding newcomers and offboarding contributors, help community managers save time while continuing to foster a collaborative environment. For instance, \citet{steinmacher2016overcoming, steinmacher2018let, steinmacher2013newcomers} highlighted that a smooth first contribution is critical for increasing engagement and retention in OSS projects. Welcome emails often have the highest open rates—up to 60\% \cite{chameleonWelcomeEmails}.

Another mechanism is to deploy pulse-check surveys that gather data on contributors' well-being and engagement levels \cite{feng2023state1}. Among the 36 OSS academic studies reviewed, 13 used surveys to understand contributors’ experiences in OSS. However, continuously deploying such surveys while addressing privacy concerns is challenging and time-consuming. Therefore, an automated survey distribution system is essential for community managers to maintain consistent communication, collect valuable feedback, and proactively identify potential reasons for attrition \cite{velykoivanenko2024designing}.

%% file: table/strategies.tex
\begin{table}
\caption{Strategies for Mitigating Challenges in OSS Retention Management}
\resizebox{5in}{!}{
\begin{tabular}{llllll}
\rowcolor[HTML]{EFEFEF} 
\textbf{ID} & \cellcolor[HTML]{EFEFEF}\textbf{Strategy}                                                                & \textbf{Description}                                                                                                                                                                      & \cellcolor[HTML]{EFEFEF}\textbf{Function}                                                    & \cellcolor[HTML]{EFEFEF}\textbf{Challenges Addressed}                                & \textbf{References}                                                                                                                                                                                                                                                                             \\
\multicolumn{6}{c}{\textbf{Community Engagement and Retention Analytics}}                                                                                                                                                                                                                                                                                                                                                                                                                                                                                                                                                                                                                                                                                                                                  \\
\rowcolor[HTML]{EFEFEF} 
S1          & \cellcolor[HTML]{EFEFEF}\begin{tabular}[c]{@{}l@{}}Track\\ Project\\ Engagement\end{tabular}             & \begin{tabular}[c]{@{}l@{}}Tracks contributor engagement \\ by analyzing activities, retention \\ trends.\end{tabular}                                                                    & \cellcolor[HTML]{EFEFEF}\begin{tabular}[c]{@{}l@{}}Diagnose \\ Disengagement\end{tabular}    & \cellcolor[HTML]{EFEFEF}\begin{tabular}[c]{@{}l@{}}C1, C2, C5,\\ C6, C7\end{tabular} & \begin{tabular}[c]{@{}l@{}}\cite{zhang2022turnover, lin2017developer, sharma2012examining}\\ \cite{qiu2019going, ait2022empirical, bao2019large}\\ \cite{schilling2014we, eluri2021predicting, zhou2012make}\end{tabular}                                                                       \\
S2          & \begin{tabular}[c]{@{}l@{}}Track\\ Newcomers\end{tabular}                                                & \begin{tabular}[c]{@{}l@{}}Tracks new contributors,\\ analyzing their participation \\ patterns and engagement levels.\end{tabular}                                                       & \begin{tabular}[c]{@{}l@{}}Diagnose \\ Attrition\end{tabular}                                & \begin{tabular}[c]{@{}l@{}}C1, C2, C4,\\ C6\end{tabular}                             & \begin{tabular}[c]{@{}l@{}}\cite{steinmacher2016overcoming, lee2018one, qiu2019going}\\ \cite{zhou2016inflow, steinmacher2015increasing, avelino2019abandonment}\\ \cite{eluri2021predicting, zhou2012make,  schilling2012will}\\ \cite{zhou2014will, steinmacher2013newcomers}\end{tabular}    \\
\rowcolor[HTML]{EFEFEF} 
S3          & \cellcolor[HTML]{EFEFEF}\begin{tabular}[c]{@{}l@{}}Track \\ Individuals \\ Contributions\end{tabular}    & \begin{tabular}[c]{@{}l@{}}Tracks individual contributor \\ activities to evaluate their level \\ of engagement, contribution \\ patterns, and impact within \\ the project.\end{tabular} & \cellcolor[HTML]{EFEFEF}\begin{tabular}[c]{@{}l@{}}Diagnose \\ Disengagement\end{tabular}    & \cellcolor[HTML]{EFEFEF}\begin{tabular}[c]{@{}l@{}}C1, C2, C5,\\ C6, C7\end{tabular} & \begin{tabular}[c]{@{}l@{}}\cite{zhou2012make, eluri2021predicting, steinmacher2015increasing}\\ \cite{zhou2014will, zhang2022turnover, scott2020productivity}\\ \cite{schilling2014we, qiu2019going, schilling2012will}\\ \cite{carillo2014s, li2021you}\end{tabular}                          \\
S4          & \begin{tabular}[c]{@{}l@{}}Predict\\ Contributor \\ Attrition\end{tabular}                               & \begin{tabular}[c]{@{}l@{}}Identifies contributors \\ who are at risk of \\ disengagement\end{tabular}                                                                                    & \begin{tabular}[c]{@{}l@{}}Diagnose \\ Attrition\end{tabular}                                & \begin{tabular}[c]{@{}l@{}}C1, C2, C3,\\ C5, C6, C9\end{tabular}                     & \begin{tabular}[c]{@{}l@{}}\cite{zhang2022turnover, lin2017developer, robinson2022two}\\ \cite{zhou2016inflow, ait2022empirical, lin2017developer}\\ \cite{low2015software, miller2019people, eluri2021predicting}\\ \cite{valiev2018ecosystem, yu2012empirical, phillips2015high}\end{tabular} \\
\rowcolor[HTML]{EFEFEF} 
S5          & \cellcolor[HTML]{EFEFEF}\begin{tabular}[c]{@{}l@{}}Understand the \\ Impact of \\ Attrition\end{tabular} & \begin{tabular}[c]{@{}l@{}}Analyzes the impact of \\ contributor turnover on \\ different project areas\end{tabular}                                                                      & \cellcolor[HTML]{EFEFEF}\begin{tabular}[c]{@{}l@{}}Diagnose\\ Project \\ Health\end{tabular} & \cellcolor[HTML]{EFEFEF}C1, C2, C3                                                   & \begin{tabular}[c]{@{}l@{}}\cite{dawn2013talent, yu2012empirical, behfar2018knowledge}\\ \cite{lin2017developer, zhang2022turnover, yu2012empirical}\\ \cite{ZHEN2009237}\end{tabular}                                                                                                          \\
S6          & \begin{tabular}[c]{@{}l@{}}Promote a \\ Welcoming\\ Environment\end{tabular}                             & \begin{tabular}[c]{@{}l@{}}Understand contributors' \\ backgrounds to foster a \\ welcoming environment.\end{tabular}                                                                     & \begin{tabular}[c]{@{}l@{}}Diagnose \\ Disengagement\end{tabular}                            & C8                                                                                   & \begin{tabular}[c]{@{}l@{}}\cite{lin2017developer, qiu2019going, trinkenreich2021please}\\ \cite{vasilescu2015gender, valiev2018ecosystem, igbaria1995impact}\\ \cite{feng2023state, rodriguez2021perceived}\\ \cite{offermann2013inclusive, sparkman2019exploring}\end{tabular}                \\
\rowcolor[HTML]{EFEFEF} 
\multicolumn{6}{c}{\cellcolor[HTML]{EFEFEF}\textbf{Project and Contributor Management}}                                                                                                                                                                                                                                                                                                                                                                                                                                                                                                                                                                                                                                                                                                                    \\
S7          & \begin{tabular}[c]{@{}l@{}}Ensure \\ Privacy \\ Practices\end{tabular}                                   & \begin{tabular}[c]{@{}l@{}}Implements data privacy to \\ protect aggregated contributor\\ information and sensitive data.\end{tabular}                                                    & Compliance                                                                                   & C10                                                                                  & \begin{tabular}[c]{@{}l@{}}\cite{qiu2019going, lin2017developer, thelwall2011privacy}\\ \cite{wong2019human, andrade2020privacy}\end{tabular}                                                                                                                                                   \\
\rowcolor[HTML]{EFEFEF} 
S8          & \cellcolor[HTML]{EFEFEF}\begin{tabular}[c]{@{}l@{}}Automated \\ Notifications\end{tabular}               & \begin{tabular}[c]{@{}l@{}}Uses automated notifications \\ to provide updates on the \\ project health.\end{tabular}                                                                      & \cellcolor[HTML]{EFEFEF}\begin{tabular}[c]{@{}l@{}}Reduce \\ Workload\end{tabular}           & \cellcolor[HTML]{EFEFEF}C2                                                           & \begin{tabular}[c]{@{}l@{}}\cite{steinmacher2016overcoming, steinmacher2018let, iqbal2014understanding}\\ \cite{lin2023automatic}\end{tabular}                                                                                                                                                  \\
S9          & \begin{tabular}[c]{@{}l@{}}Utilize Tools for\\ Contributor \\ Engagement\end{tabular}                    & \begin{tabular}[c]{@{}l@{}}Employs engagement tools \\ to enhance contributor \\ interactions\end{tabular}                                                                                & \begin{tabular}[c]{@{}l@{}}Reduce \\ Workload\end{tabular}                                   & C2, C4                                                                               & \begin{tabular}[c]{@{}l@{}}\cite{steinmacher2018let, carillo2014s, steinmacher2016overcoming}\\ \cite{iqbal2014understanding, steinmacher2013newcomers}\end{tabular}                                                                                                                           
\end{tabular}}
\label{tab:strategies}
\end{table}

%% file: section/sec6_FGD.tex
\section{Prototype Design (\faUsers)} 
\label{sec:stratagies-ii}

\begin{figure}[!bt]
    \centering
    \includegraphics[width = \textwidth]{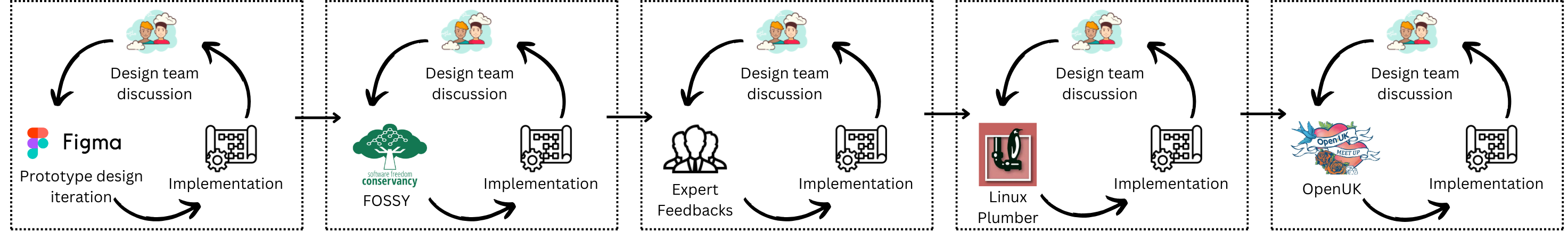}
    \caption{User-centered design process illustrating iterative design and implementation cycles, incorporating multiple focus group discussions with over 100 OSS stakeholders.}
    \label{fig:FGD}
    \vspace{5mm}
\end{figure}

Following the DSR paradigm, we decided to evaluate the efficacy of the above strategies by operationalizing them in a web-based prototype, enabling an in situ evaluation. The prototype acts as a vehicle for intervention and knowledge generation, as DSR emphasizes validating technological rules through real-world instantiation \cite{runeson2020design, engstrom2020software}. Without a functional prototype reflecting real-life data from their own projects, participants would be unable to assess the strategies’ effectiveness. Survey or interview-based evaluation requires recall of past experiences, whereas prototypes anchor feedback in interactive experiences, enhancing the consistency and reliability of responses \cite{furnham1986response}. We followed a user-centric approach when implementing the strategies in our prototype by conducting iterative focus group discussions with OSS stakeholders (OSS community managers and maintainers) (Figure~\ref{fig:FGD}). These discussions helped: (1) bridge user expectations by clarifying their needs and identifying obstacles in integrating the strategies into real-world context \cite{simko2021would}; (2) receive continuous feedback and make iterative modifications based on participants' interactions with the prototype \cite{lam2011empirical, zhuang2022framework}; and (3) validate and prioritize features, ensuring that the prototype aligned with both technical feasibility and user expectations \cite{da2011user}.

This iterative refinement process contributes to the rigor of our study, as recommended in the DSR paradigm, by gradually expanding the scope and realism of validation while managing the risks associated with deploying interventions in operational OSS contexts \cite{runeson2020design, engstrom2020software}. Our iterative, context-sensitive approach, rooted in stakeholder engagement and practical feasibility, helps mitigate potential disruption to ongoing OSS activities while ensuring that our strategies reflect usable and reliable design knowledge prior to in situ evaluation.

\subsection{Prototype Walkthrough}

\begin{figure}[!htb]
    \centering
    \includegraphics[width = 0.9\textwidth]{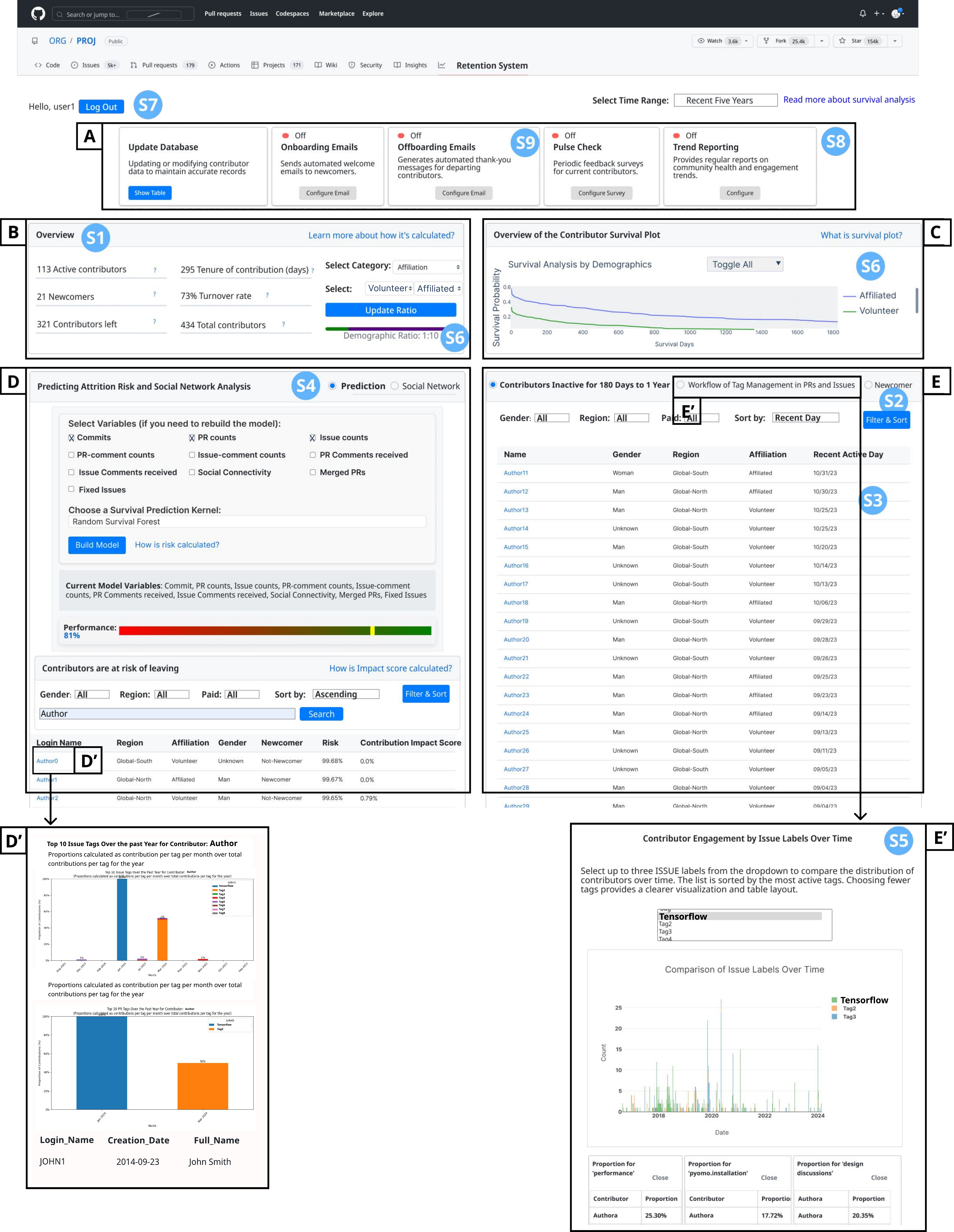}
    \caption{\protect\overviewLabelBig{S1}–\protect\overviewLabelBig{S9} represent strategy indices, as defined in Table~\ref{tab:strategies}. Sections A–D illustrate different components of the prototype, including automation setup (A), project retention overview (B), survival analysis (C), contributor attrition analysis (D) and inactive/newcomer/Tag management (E). Additional detailed views are indicated as D’ and E’, corresponding to contributor-specific data and tag management. See the supplementary materials for the walk-through video \cite{suppl}.}

    \label{fig:overview}
\end{figure}

Before detailing how we employed the user-centric approach to implement these strategies, we begin by presenting a walkthrough of the finalized prototype (Figure \ref{fig:overview}) to anchor and contextualize our subsequent discussion.

Let's take the example of Tyler, a community manager of an OSS project using our prototype. Upon logging into the system, features in Section A provide automated support for engagement and trend reports; using this, Tyler sets up onboarding emails to welcome new contributors and schedules offboarding emails to stay engaged with departing contributors.

Tyler next explores features in Section B, which offers an overview of the project’s retention metrics. This section displays statistics such as the number of active contributors, the number of newcomers, and the project’s turnover rate. Tyler also checks the distribution of contributors across demographic groups—specifically comparing affiliated contributors to volunteer contributors. Noticing that the proportion of affiliated contributors (represented by the purple bar) is significantly wider than that of volunteers (green bar), Tyler suspects a potential decline in volunteer retention. To investigate further, Tyler compares the retention rates between these two groups.

Shifting the focus to Section C, Tyler checks the survival analysis by comparing affiliated and volunteer contributors, observing that the survival rate for volunteers (green line) is lower than that for affiliated contributors (purple line). The analysis reveals that less than 20\% of volunteer contributors remain active after one year, prompting Tyler to identify individuals at the highest risk of leaving the project.

To continue this investigation, Tyler then moves to Section D to create a prediction model in investigating the high-risk contributors who are leaving, using contribution activities such as commits, pull requests, and issue tracking; they build a Random Survival Forest model \cite{ishwaran2008random} to predict contributors at risk of disengagement. Upon identifying a contributor with the highest risk score, Tyler clicks on their profile, triggering a detailed popup (D’) that outlines the contributor's activity history. They notice this contributor has been actively involved with pull requests tagged under [TensorFlow API].

To assess the potential impact of this contributor leaving the project, Tyler proceeds to the PR Tag Management System (Section E’) and selects the [TensorFlow API] tag; they verify that this individual is not the most critical contributor for that area. Therefore, there is no high risk for the projects. Tyler repeats this evaluation for other high-risk contributors.

Lastly, to ensure that no contributors are overlooked, still within Section E, Tyler switches to the Newcomer Table tab and Inactive Contributor tab to track recent newcomers and individuals who have been inactive for over six months.

For additional prototype features and detailed technical specifications, please refer to the supplementary documents \cite{suppl}.

\subsection{User-centric Design Iteration}

In this section, we describe how we operationalized different strategies in our prototypes discussed in Section \ref{sec:stratagies}. We focus on identifying the information needed for each strategy, the data sources leveraged, and the methodological decisions of implementation.

\subsubsection{Step 1: Prototyping Strategies:} The first step in operationalizing the strategies was to prototype each strategy and identify what information was needed.  Our research team held weekly meetings (throughout one month) to discuss how to prototype each strategy, focusing on identifying relevant data sources and determining the appropriate methods for implementation, which we describe below.

\textbf{Retention Metrics to Understand Engagement:} We decided to use \textit{retention metrics} that have been discussed in existing OSS studies to operationalize \textsc{S1. Track Project Engagement} and \textsc{S2. Track Newcomers}, including the number of active contributors \cite{zhou2016inflow}, newcomers \cite{zhou2016inflow, steinmacher2016overcoming, steinmacher2015increasing, steinmacher2018let, steinmacher2013newcomers, yu2012empirical}, number of contributors who have left \cite{zhou2016inflow, zhou2012make, avelino2019abandonment, zhang2022turnover, iaffaldano2019developers, schilling2014we, vasilescu2015gender}, total contributors \cite{zhou2016inflow, avelino2019abandonment, zhou2012make, midha2007retention}, turnover rate \cite{zhou2016inflow, qiu2019going, avelino2019abandonment, zhang2022turnover, lin2017developer, yu2012empirical, vasilescu2015gender}, and the average tenure of contribution (in days) \cite{zhou2016inflow, qiu2019going, zhou2012make, zhang2022turnover, schilling2012will, schilling2014we, tsay2014influence}.  Furthermore, it is also important to provide community managers with a list of newcomers. By identifying newcomers early, community managers can proactively offer mentorship or resources that improve their experience, ultimately increasing retention and contributing to a sustainable contributor pipeline \cite{feng2022case, steinmacher2014attracting, steinmacher2016overcoming}.

\textbf{Contributor Engagement and Attrition Risks:} 
Community managers need detailed information to \textsc{S3. track individual contributors}, such as each contributor’s contribution history, specificity (the part of the project they have worked on), and social network. Access to this information helps community managers better understand contributors' engagement and identify potential factors that may influence their participation \cite{cheng2019activity}.

In addition to static engagement metrics, proactive monitoring of contributors' activity patterns can provide community managers with early indicators to \textsc{S4. understand contributor attrition}. Studies in OSS have utilized statistical models to predict the likelihood of individual contributors leaving a project \cite{eluri2021predicting, eluri2021predicting, schilling2012will, giovanini2021leveraging, bao2019large}.

Once community managers acquire proactive information about declining contributor activity and engagement, it's important to \textsc{S5. understand the impact of attrition} on the project. This involves assessing the contributor's past contributions, determining the project areas that might be affected by their departure, and identifying other team members who can assume those responsibilities. One approach is to leverage the existing tagging systems widely used in issue-tracking platforms \cite{arya2019analysis}. For instance, issue tags can quickly indicate categories such as bug, feature, and optimization, helping contributors identify areas of interest or expertise \cite{kalliamvakou2017makes}. Similarly, in some communities, pull request tags can indicate the progress or priority of changes \cite{kalliamvakou2017makes}. Therefore, by analyzing the distribution of contributions across these tags and identifying the contributors associated with them, one can gain insights into who is working on specific areas \cite{vasilescu2015quality}. If a contributor who has been heavily involved in a particular tag stops contributing, community managers can identify the gap to find another contributor to take on the responsibility.

To \textsc{S6. promote a welcoming environment}, it is important for community managers to understand if certain demographic groups have a higher risk of attrition \cite{qiu2019going, avelino2019abandonment, lin2017developer, vasilescu2015gender}; for instance, monitoring retention rates between affiliated and non-affiliated contributors can reveal disengagement patterns \cite{guizani2023rules}. Existing studies have used the Kaplan-Meier survival analysis estimate to compare demographic groups and identify potential retention imbalances within the community \cite{qiu2019going}. In our prototype, we chose to use the commonly studied demographic lenses in OSS research: newcomer status, gender, region, and affiliation \cite{qiu2019going, avelino2019abandonment, lin2017developer, vasilescu2015gender, feng2023state1}.

\textbf{Access Controls to Maintain Privacy Practice:} For all of the information and aggregated data we have discussed above, access control mechanisms to \textsc{S7. ensure privacy practices}, such as secure login systems, are essential in protecting sensitive data within OSS projects to reduce the risk of unauthorized exposure  \cite{payne2002security}. Features such as a login page that only allows community managers to sign up -- with sign-up requests being sent to a designated administrator for approval -- help ensure that sensitive information is kept private to the fullest extent possible.

\textbf{Automated Approaches to Community Health Monitoring and Contributor Engagement:} \textsc{S8. Automated notification to monitor community health} is essential to help community managers with their workload. In our prototype, we incorporated a feature that allows community managers to schedule notifications for project reports that include data such as the number of newcomers who have joined or contributors who may be leaving. This approach ensures that managers receive information they consider relevant in a convenient manner.

Regularly engaging with contributors is vital for addressing disengagement early \cite{dabbish2012social, steinmacher2014attracting}. However, managing high volumes of routine communication can be time-consuming and may lead to burnout. To mitigate this, we propose to \textsc{S9. utilize tools for contributor engagement}. For example, automating the sending of welcoming emails to onboard newcomers, expressing gratitude to departing contributors, and conducting wellness surveys via email can foster a positive community environment. These automated interactions help maintain consistent communication and support within the community.

\subsubsection{Step 2: Initial Prototype:}
We designed the initial prototype by following the design principles outlined in \citet{few2006information} to display the necessary information from the previous step. Three researchers summarized these design principles, ensuring that each design element chosen for implementation was grounded in these established guidelines. We began with low-fidelity prototyping and conducted two sandbox sessions to finalize the low-fidelity prototypes. We then used Figma\footnote{\url{https://www.figma.com/prototyping/}} to design the prototype with interactions. The detailed design guidance we followed can be found in supplementary documentation \cite{suppl}.

\subsubsection{Step 3: Collecting feedback from stakeholders:} The next step in our design approach was to gather stakeholder feedback, allowing stakeholders to freely express their ideas and provide insights on how users might interact with the tool \cite{bodker2022participatory}. As shown in Figure \ref{fig:FGD}, we collected four rounds of feedback, including three focus group discussion sessions with OSS conference participants and 10 one-on-one consultation sessions with OSS community managers.

\textbf{Stage I. Initial Feedback from Focus Group Discussion I:} We had our first focus group discussion at the FOSSY Conference 2023 \cite{FOSSY}. We used a Figma high-fidelity prototype for this session instead of an implemented version for flexibility in quick iterative improvements.

The focus group discussion session was about an hour long, and we spent the first 10 minutes showcasing the prototype's features and functionalities, while the remaining time was allocated for audience feedback. The first and last authors led the group discussion. There were 20 participants from whom we collected various suggestions, which ranged from topics such as UX design to backend work. One of the main pieces of feedback we received was about predicting the likelihood of a contributor leaving the project. Our audience suggested that the variables used in predictions should not be fixed, as different projects have different engagement patterns. They emphasized that the users should decide the metrics, as different communities can have unique characteristics and needs.

\textit{Discussion and Improvement:} After the focus group discussion, the research team convened to discuss and iterate over these feature requests and implementation through weekly meetings spanning one month.  To better incorporate the main suggestion from the audience, we also agreed that when predicting the likelihood of individuals leaving the project, we should not be limited to a particular statistical model (regression). Thus, we decided to implement all three standard survival models: Cox regression \cite{cole1993cox}, neural network \cite{faraggi1995neural}, and random forest \cite{ishwaran2008random}. This approach gives users more flexibility to fit different variables and select models based on their preferences and needs. Additionally, we provide an indicator for showing the performance of each model to help them select the most suitable model (for their metrics of interest). This ensures that users can make data-driven decisions when interpreting and applying their projects' results.

Another requested feature was to list inactive contributors \cite{CHAOSS}, defined as those who have paused their contributions for more than six months but less than a year. This report allows community managers to identify and target these contributors with tailored re-engagement strategies.

\textbf{Stage II. Initial Implementation and Consultations with Kubernetes and Flutter:} In this stage, we implemented the first version of the prototype. The initial implementation closely followed the high-fidelity prototype. We chose to use the Flask framework of the Python language \cite{aslam2015efficient} for implementation, as it is a low-cost, efficient, and quick approach.

Instead of placeholders for data usage, we used actual OSS project data for the implementation, which also helped test the prototype's scalability and usability. We chose Flutter\footnote{Flutter GitHub Repository. Available at: \url{https://github.com/flutter/flutter}} as a use case because (1) it has a large community with more than 500 contributors and over 10 years of history and (2) is very active, with more than 10 issues/pull requests updated every day. Additionally, we had access to talk to the Flutter community managers to collect feedback.

All features were implemented using CSS, HTML, and JavaScript. REST APIs were leveraged to pull data from Flutter's GitHub repository. For this proof of concept implementation, we chose the simplest inference tools for identifying contributor demographic information.
We inferred contributor affiliations based on email domains (e.g., google.com vs. gmail.com). Non-corporate email domains were classified as unknown. To infer the gender and region of contributors, we used the Namsor API \cite{sebo2021performance}, a name inference tool that estimates the gender of an individual on a probability scale from -1 to +1, by incorporating the individual's full name and geographic information. However, relying solely on geographic data from GitHub profiles reduces precision \cite{vasilescu2014gender}, mainly when contributors' GitHub locations differ from the geographic origin of their names. To address this, we implemented a two-step process: first, we used the Namsor API to predict the geographic origin of contributors' names; then, we used both the predicted origin and the full names to infer contributors' gender \cite{guizani2025community}. To minimize errors, we applied a probability threshold, filtering out predictions with confidence levels below 90\%, in line with best practices from prior research \cite{carsenat2019inferring, sebo2021performance}. 
For implementing the survival analysis for Flutter, we conducted preliminary tests to validate the reliability of our approach as detailed in \citet{feng2023state}. This prior work provides a foundation for incorporating survival analysis into our prototype.

\textit{Consultations with Expert (Flutter):} After implementing the system, we had a two-hour-long meeting with the project manager of Flutter to walk him through the system and its features. He validated the need for these strategies when it came to community management and provided valuable feedback. For example, he suggested creating a contribution impact score to quantify and compare the impact of individual contributors’ efforts in an aggregated form.

\textit{Discussion and Improvement:} We addressed this primary feedback through the implementation of a ``contribution impact score'', which calculates a contributor's score by considering their number of contributions in proportion to a baseline score configured as the highest number of contributions made by a contributor. For example, if the top contributor has 50 commits, another contributor with 40 commits would receive a score of 0.8. In our case, we treated all types of contributions as equal as a proof of concept. Future iterations can refine this strategy by assigning weights to different contribution types for a more nuanced assessment tailored to the specific needs of the OSS community.

\textbf{Consultations with Experts (community leaders from CNCF):} We collected the next round of feedback with the modified prototype. We attended KubeCon 2023 \cite{KubeCon2024} and invited 10 community leaders from different communities to review our (implemented) strategies.

The sessions were informal, with each lasting 15-20 minutes. We used the first three minutes to introduce the strategies through a walkthrough of the prototype and then let the project leaders freely explore the system to provide feedback.  
We received comments on what data sources to use as well as the look-and-feel of the prototype itself. One community leader suggested that strategies for engagement need to cover the entire community, which includes more than just the main Kubernetes repository, as contributors in their community may contribute to different side repositories depending on their project goals. Another project leader had concerns about collecting demographic information, particularly because of privacy issues, and how to ethically ask people to identify themselves.

\textit{Discussion and Improvement:} Similar to the previous steps, we revisited and discussed the feedback with our design team, refining the prototype accordingly. However, we were unable to find a suitable solution regarding the privacy concerns related to the contributor's demographic information. As a result, we decided to hold another focus group discussion (see next stage) to gather additional feedback from a broader range of participants.

\textbf{Stage III. Iterative Refinement from Focus Group Discussion II:} 
We hosted a focus group discussion session at the Linux Plumbers Conference \cite{LPC2025}, where the majority of participants were OSS maintainers, allowing us to get their feedback. During the discussion, we mainly focused on how to help community managers understand the contributors' background and privacy issues (one of the sticking points from the prior stage). We presented our approach to infer contributors' identity by using research tools to infer their gender and region. Although participants had concerns about approaches to infer contributors' identities, this feature was also highly voted by the OSS community, so we could not simply drop it. Some participants mentioned that if we use inference tools for gender identity and a contributor sees it, it could offend the contributor if their gender is misclassified. Additionally, some participants suggested we could ask contributors to self-report their identities, but others raised concerns about newcomers having to self-report their identity when joining a project, as it may be concerning for those who do not want to disclose this information or are unsure about their long-term commitment.

\textit{Discussion and Improvement:} We updated our prototype design such that demographic information remains private behind login pages and administrators must approve access to this information. Additionally, maintainers felt that such a system was not meant for the general public (where there can be more negative repercussions of revealing demographics), but could be used by community managers to manage retention. We also gave project leaders the option to create wellness check surveys so that contributors can self-report their demographics if they feel comfortable. Additionally, we provided a mechanism to allow managers to review and correct the inference tool’s outcomes to improve accuracy and give the community more flexibility.

We acknowledge that we implemented strategy (S6) by using automated inference of demographic information, which may be unsuitable for real-world applications. Future implementations of this strategy need to be designed for transparency and ethical data handling of sensitive data in alignment with varying community norms, data protection regulations, and cultural contexts.

\textbf{Stage IV. Final Feedback from Focus Group Discussion III:}
With the updated version of the prototype, we conducted another focus group discussion at OPENUK 2024 \cite{OpenUK}, with over 30 remote participants for a 30-minute session. Ten minutes were devoted to introducing the strategies through a prototype walkthrough, followed by 20 minutes of open discussion primarily focused on clarifying the prototype and its features. We received no further comments or suggestions; several attendees expressed interest in using the prototype, and four participants later emailed the speaker to inquire about adopting it for their communities. This feedback indicated that the prototype was ready for in situ evaluation (See the supplementary documentation \cite{suppl} for a prototype walk-through video).

%% file: section/sec7_User_evaluation.tex
\section{In Situ User Evaluation \faUserEdit}
\label{sec:evaluation}

To enhance the rigor of our study and evaluate the strategies in real-world settings, we conducted an in situ user evaluation, using the prototype features as a proxy to assess the effectiveness of the retention strategies. In situ evaluation refers to assessing an artifact within the actual environment in which it is intended to be used, offering ecological validity by capturing realistic workflows, stakeholder roles, and decision-making contexts \cite{consolvo2007conducting}. As demonstrated in \citet{lechelt2021evalme}, in situ approaches allow participants to reflect on and interact with the system “in the moment,” providing authentic, temporally grounded feedback that might otherwise be lost in post-hoc reflections or surveys. We performed this evaluation by applying the prototype to two large OSS communities—Pyomo/pyomo \cite{pyomo2024} and Microsoft/DeepSpeed \cite{deepspeed2024}. Both projects have more than 100 contributors.

We selected the projects to ensure they differed in their governance structures and application domains to understand the generalizability and practicality of the implemented strategies. Pyomo is a Python-based scientific software package widely used in academic settings, offering optimization tools for modeling and analysis. Over the past seven years, it has been developed and maintained by a community of researchers and practitioners in their spare time. DeepSpeed is a deep-learning optimization library for PyTorch, sponsored by Microsoft and commonly adopted in the software industry. It has nearly five years of history and a more corporate governance model.

\subsection{Method}

\textbf{Evaluation Approach.} We adopted a cross-sectional evaluation of the project primarily to obtain immediate, actionable insights about the effectiveness of the implemented strategies. Cross-sectional evaluation designs have been widely validated for capturing current user behaviors and adoption potential \cite{wang2020cross, kujala2019cross}, making them well-suited for rapid assessment. By contrast, longitudinal approaches (e.g., diary studies) would be complicated by the inherently dynamic nature of OSS projects. For example, new package releases or onboarding programs could introduce major shifts in retention. To account for differences in perspectives and needs of individual managers, we invited all the governance members (project community managers) from both Pyomo and DeepSpeed to participate in the study.

\textbf{Evaluation Questions.} 
As we used a web-based prototype to implement the strategies, we leveraged the evaluation framework proposed by \citet{lam2011empirical} for visualization tools. These include: 
(1) What features are useful? (2) What features are missing? (3) How can features be reworked to improve the supported work processes? (4) Are there limitations to the current prototype that would hinder its adoption? (5) Is the prototype understandable, and can it be learned?

We group these questions into two top-level evaluation questions:

\textit{EQ1. How does the prototype help project managers mitigate community management challenges?} This covers sub-questions (1), (2), and (5), and focuses on the prototype’s usefulness, its ease of learning, and any missing features. It prompts participants to reflect on how the prototype supports their retention management needs. 

\textit{EQ2. How can the prototype complement project managers’ current retention management approaches?} This aligns with sub-questions (3) and (4), assessing how the prototype integrates into existing workflows and identifying potential barriers to adoption.

\textbf{Participant recruitment.} We first reached out to the project managers of Pymo and DeepSpeed who had expressed interest in evaluating our prototype. These managers then helped us recruit the remaining project manager team members. We recruited eight participants in total: four from Pyomo and four from DeepSpeed. Table \ref{tab:user_study_demo} shows participants' demographics.

\input{table/user_study_demo}

The study design was approved by the university’s IRB, which determined that the study involved minimal risk. We compensated our participants with a \$50 gift card for their time. The study was conducted in-person (n=3) or remotely (n=5) via Zoom, depending on the participants' preferences. All interactions with the prototype were browser-based.

\textbf{Evaluation protocol.} The study began with an introduction to the project, which described the structure of the study, followed by three demographic questions: gender, OSS seniority, and project leadership status. Participants then viewed a six-minute tutorial video. The video was designed to ensure that information about the prototype was presented consistently to all participants. The participants were then free to explore the prototype while thinking aloud about how they would use it to manage their community, highlighting any features and information they found interesting or that needed improvement. To ensure that participants understood the think-aloud process, we prepared a practice session to walk them through the think-aloud protocol.

After the exploration, participants completed two sets of post-study questionnaires. The first set of questions aims to understand participants' agreement on \textit{``how each implemented strategy within the prototype helped to mitigate retention challenges''}. The second set explored \textit{``how the prototype could complement or enhance participants’ existing community management strategies''}. Additionally, we included questions such as whether they would be willing to purchase it, recommend it to friends, and continue to use it. Lastly, we asked open-ended questions, focusing on which features they liked or disliked and any other recommendations they had. See the supplementary documentation \cite{suppl} for details about evaluation design.

Before conducting the study with participants, we piloted it with two graduate research assistants experienced in OSS research. Based on their feedback, we revised the study design, adding a think-aloud practice to support participants unfamiliar with the method. To gather further input, we presented our study design to a graduate HCI design class at Oregon State University, involving 30 graduate students. Feedback from the students and the professor led to additional revisions, such as inviting a native English speaker to record the tutorial video to minimize language barriers. Finally, we conducted two more pilot studies with OSS practitioners, iterating until no further feedback or suggestions were provided. Refer to the supplementary materials \cite{suppl} for details about the study, tutorial videos, and study scripts.

\textbf{Data analysis.} We began by transcribing the video and audio recordings. We used qualitative analysis to answer the evaluation questions by analyzing participants' user experiences, their think-aloud responses, and answers to post-study open-ended questions. Two authors independently coded the four interview transcripts to identify initial themes. We then met to compare and discuss codes using the constant comparison method, refining them into a preliminary codebook by merging overlapping codes and creating new ones where necessary. The same two authors independently coded the remaining interview transcripts using this codebook. After completing the coding, they met again to review the code, negotiate differences, ensure consistency, and resolve discrepancies. We also quantitatively analyzed the post-study questionnaire data to complement our qualitative findings.

\subsection{Results}

\input{table/user_study}

\textbf{EQ1. Helpfulness in Mitigating Challenges:} Column one in Table \ref{tab:user_study} presents participant responses on how our prototype could mitigate community management challenges. Among the eight participants, there was strong agreement across all questions, with no more than one disagreement for any challenge across the two projects.

Community managers found the prototype helpful in \textsc{managing overwhelming responsibilities} [PY3, PY4, D1–D4] \footnote{PY: Pyomo participant; D: DeepSpeed participant.}. \textit{``I think the most helpful features are to understand contributions—specifically, the contributor activity section, which shows the top contributors for different areas. That is very important, along with the at-risk information, as both help me plan the project's features''} [D2].

Participants emphasized the prototype's helpfulness in \textsc{managing retention}. \textit{``Seeing the overall retention trend. That's just interesting''} [PY1]. \textit{``I know people leaving, how we can do with this limited bandwidth. I'm always feeling guilty to not give praise. Early sign. When I notice it is too late''} [D1]. Similarly, a participant noted the difficulty of tracking turnover manually, explaining, \textit{``The turnover rate, I guess... Never thought of our repo that way... It's not easy to track when someone's gone... seeing it in a number is more straightforward”} [D2]. \textit{``Our biggest issue with retention is just actually being able to get to and solve issues. If it helped us do that in a better way, especially with our top contributors, solving their issues, you know, giving that priority. That would be a big help''} [D3].

The prototype received unanimous agreement for its helpfulness in \textsc{engaging contributors} [PY1–PY4, D1–D4]. Participants emphasized the importance of understanding and interacting with contributors, particularly less visible ones. \textit{``Interacting with contributors... we get asked a lot who our users are. We actually really don’t know... but the quiet people using Pyomo, we have no idea''} [PY4]. D4 mentioned the potential to foster engagement: \textit{``You can send the email... help contributors feel welcome, especially people [that] are not affiliated, but you encourage them to contribute''}. Additionally, the idea of automated features for personalized welcome emails resonated with PY2: \textit{``Having an automated system to do kind of a personalized welcome, I think, is pretty neat... I’m not aware of an easy way to do that''}.

Another unanimous agreement was on the prototype's helpfulness in \textsc{tracking contributors' data} [PY2–PY4, D1–D4]. \textit{``If you need to learn about somebody who has done something, you have to go to GitHub and check... There’s no way... aggregated, and all the contribution data of one person''} [D2]. Similarly, PY4 mentioned, \textit{``Aggregating the different types of contributions... I actually really like that. I like what you're pulling out about... whether certain sub-packages and those sorts of things are getting support''}.

Community managers found the prototype particularly helpful for \textsc{recognizing hard-to-track contributions} that are not easily accessible through GitHub's interface, which helped them better understand the extent of each contributor's engagement [PY1–PY4, D1–D4].  \textit{``This is the most helpful for me because it gives me a chance to see people I don’t recognize and remember why they contributed''} [PY1]. \textit{``I think there are a lot of contributions, particularly in, you know, contributing ideas through issue comments or contributing PR reviews and things like that. So just contributions to keeping the repo healthy''} [PY2].

Another aspect where the prototype demonstrated its helpfulness was in \textsc{fostering inclusive collaboration} [PY2, PY4, D1, D3, D4]. PY1 mentioned \textit{``I like that idea a lot... In a community like Pyomo or Spack, which is huge, you can toggle different demographics to compare with the overall. If a contributor is left out, it can help identify if they need more support''}. Similarly, PY2 mentioned \textit{``Assessing the demographics of those who may disengage... I haven't seen any other tools for gathering GitHub statistics like this''} [PY2].

However, we observed some reasons for disagreement in certain areas. Regarding \textsc{maintaining privacy}, the responses were mostly neutral due to differing work environments. \textit{``Maintain privacy when aggregating contributors' data... I'm going to put this as neutral because, you know, coming from a national lab viewpoint, we have some strict ideas and policies around privacy''} [PY2].

\textbf{EQ2. Complementing Community Managers' Current Management Approach:} Column two in Table \ref{tab:user_study} shows participants' agreement on how the prototype complements their current retention management strategies. The majority of participants agreed that the prototype effectively complements their current strategies.

Participants agreed that the prototype could assist in tracking contributors' activities and engagement [PY1–PY3, D1–D4] and enhance interactions with contributors [PY2, PY3, D1–D4]. \textit{``I think these statistics are interesting... seeing changes in activity could be interesting and useful. I also think engaging with contributors is by far probably the most useful feature''} [PY2]. One community manager from DeepSpeed was surprised by the data provided, \textit{``Whoa, whoa, [...] active contributors''} [D1], and [D4] noted, \textit{``This could be super helpful for understanding which contributors may engage''}.

In addition to tracking overall engagement, participants unanimously agreed on the value of tracking newcomers [PY1–PY4, D1–D4]. \textit{``Tracking how many newcomers... must be useful''} [D1]. \textit{``I really like that in particular. The idea of being able to see the newcomers... especially in a community like Pyomo, where a lot of us are paid to do it''} [PY1].

However, there were a few aspects where we observed one instance of disagreement, particularly around data privacy and contribution attrition. The reasons for this stemmed from differing management and OSS philosophies. PY1 mentioned that \textit{``Maintaining privacy for contributors’ data... we don’t really care, so probably not just for us''}. \textit{``Privacy part... I mean, it's all just kind of public data, in a way, anyways''} [D4]. Similarly, in discussing contribution attrition, \textit{``For us... that's actually probably not the most important thing. I think there are a lot of people we know we're going to lose anyway''} [PY4].

Lastly, column three of Table \ref{tab:user_study} shows participants responses for future adoption; we asked participants three questions regarding their willingness to continue using the tool, recommend it to other managers, and purchase it. All participants indicated they would recommend our prototype to other managers. \textit{``I’m gonna say, I would definitely recommend it''} [PY2]. Other participants highlighted how it could help balance workloads: \textit{``You know, we only have limited space to run PRs, so a lot of our stuff gets backed up''} [D3] and \textit{``[it] could be helpful in balancing the workload across different projects''} [PY2]. Regarding the willingness to purchase the tool, participants gave careful consideration, mentioning that any purchase decision would need to go through community management meetings: \textit{``So the reason I'll say `maybe' on the willingness to purchase is because we don't have a lot of direct money''} [PY1].

%% file: table/user_study_demo.tex
\begin{table}[!tbp]
\caption{Demographic information of interview participants.}
\resizebox{3.5in}{!}{
\begin{tabular}{lllll}
\rowcolor[HTML]{EFEFEF} 
\textbf{ID} & \textbf{Community} & \textbf{Gender} & \textbf{OSS Experiences} & \textbf{OSS Management Experience} \\
PY1          & Pyomo              & Man             & 3 to 5 years             & 3 to 5 years                       \\
\rowcolor[HTML]{EFEFEF} 
PY2          & Pyomo              & Woman           & 3 to 5 years             & 3 to 5 years                       \\
PY3          & Pyomo              & Woman           & 6 to 10 years            & 3 to 5 years                       \\
\rowcolor[HTML]{EFEFEF} 
PY4          & Pyomo              & Woman           & > 10 years               & 3 to 5 years                       \\
D1          & DeepSpeed          & Man             & 3 to 5 years             & 3 to 5 years                       \\
\rowcolor[HTML]{EFEFEF} 
D2          & DeepSpeed          & Man             & 3 to 5 years             & 3 to 5 years                       \\
D3          & DeepSpeed          & Man             & 3 to 5 years             & 1 to 2 years                       \\
\rowcolor[HTML]{EFEFEF} 
D4          & DeepSpeed          & Man             & 1 to 2 years             & 1 to 2 years                      
\end{tabular}}
\label{tab:user_study_demo}
\end{table}

%% file: table/user_study.tex
\begin{table}[!tbp]
\caption{The table presents agreement levels (N=8) for identified challenges, proposed strategies, and future adoption intentions.}
\resizebox{5.5in}{!}{%
\begin{tabular}{llcllcll}
\multicolumn{1}{c}{\cellcolor[HTML]{EFEFEF}\textbf{Mitigate Challenge?}}                                                               & \multicolumn{1}{c}{\cellcolor[HTML]{EFEFEF}\textbf{Agreement (N=8)}} &                     & \multicolumn{1}{c}{\cellcolor[HTML]{EFEFEF}\textbf{Complement Strategy?}}                                                             & \multicolumn{1}{c}{\cellcolor[HTML]{EFEFEF}\textbf{Agreement (N=8)}} &                     & \multicolumn{1}{c}{\cellcolor[HTML]{EFEFEF}\textbf{Future Adoption?}}                      & \multicolumn{1}{c}{\cellcolor[HTML]{EFEFEF}\textbf{Agreement (N=8)}} \\
\begin{tabular}[c]{@{}l@{}}Overwhelming\\ Responsibilities\end{tabular}                                                       & \fivepointlikertt{3}{3}{1}{0}{1}                                      &                     & \begin{tabular}[c]{@{}l@{}}Track\\ Project\\ Engagement\end{tabular}                                                      & \fivepointlikertt{5}{1}{2}{0}{0}                                       &                     & \begin{tabular}[c]{@{}l@{}}Will you \\ recommend?\end{tabular}                   & \fivepointlikertt{8}{0}{0}{0}{0}                                         \\
\cellcolor[HTML]{EFEFEF}\begin{tabular}[c]{@{}l@{}}Challenges of\\ Managing\\ Turnover\end{tabular}                           & \cellcolor[HTML]{EFEFEF}\fivepointlikertt{3}{3}{1}{0}{1}              &                     & \cellcolor[HTML]{EFEFEF}\begin{tabular}[c]{@{}l@{}}Track\\ Newcomers\end{tabular}                                         & \cellcolor[HTML]{EFEFEF}\fivepointlikertt{7}{1}{0}{0}{0}                &                     & \cellcolor[HTML]{EFEFEF}\begin{tabular}[c]{@{}l@{}}Will you \\ use?\end{tabular} & \cellcolor[HTML]{EFEFEF}\fivepointlikertt{7}{1}{0}{0}{0}                \\
\begin{tabular}[c]{@{}l@{}}Time-consuming to \\ Engage and Grow \\ the Contributor Base\end{tabular}                          & \fivepointlikertt{4}{3}{1}{0}{0}                                       &                     & \begin{tabular}[c]{@{}l@{}}Track\\ Individuals'\\ Contributions\end{tabular}                                               & \fivepointlikertt{3}{3}{2}{0}{0}                                       &                     & \begin{tabular}[c]{@{}l@{}}Will you \\ purchase?\end{tabular}                    & \fivepointlikertt{1}{5}{0}{2}{0}                                       \\
\cellcolor[HTML]{EFEFEF}\begin{tabular}[c]{@{}l@{}}Challenges in \\ Tracking \\ Project Data\end{tabular}                     & \cellcolor[HTML]{EFEFEF}\fivepointlikertt{3}{2}{2}{1}{0}             &                     & \cellcolor[HTML]{EFEFEF}\begin{tabular}[c]{@{}l@{}}Predict\\ Contributor\\ Attrition\end{tabular}                      & \cellcolor[HTML]{EFEFEF}\fivepointlikertt{3}{2}{2}{1}{0}              &                     & \cellcolor[HTML]{EFEFEF}                                                         & \cellcolor[HTML]{EFEFEF}                                             \\
\begin{tabular}[c]{@{}l@{}}Challenges in \\ Tracking \\ Contributor Data\end{tabular}                                         & \fivepointlikertt{4}{3}{1}{0}{0}                                       &                     & \begin{tabular}[c]{@{}l@{}}Understand the\\ Impact of\\ Attrition\end{tabular}                                            & \fivepointlikertt{4}{2}{1}{1}{0}                                      &                     &                                                                                  &                                                                      \\
\cellcolor[HTML]{EFEFEF}\begin{tabular}[c]{@{}l@{}}Challenges in\\ Acknowledging\\ Hard-to-Track\\ Contributions\end{tabular} & \cellcolor[HTML]{EFEFEF}\fivepointlikertt{5}{3}{0}{0}{0}                &                     & \cellcolor[HTML]{EFEFEF}\begin{tabular}[c]{@{}l@{}}Promote a\\ Welcoming\\ Environment\end{tabular}                        & \cellcolor[HTML]{EFEFEF}\fivepointlikertt{1}{5}{1}{1}{0}              &                     & \cellcolor[HTML]{EFEFEF}                                                         & \cellcolor[HTML]{EFEFEF}                                             \\
\begin{tabular}[c]{@{}l@{}}Challenges in\\ Fostering Inclusive\\ Collaboration\end{tabular}                                   & \fivepointlikertt{2}{3}{3}{0}{0}                                       &                     & \begin{tabular}[c]{@{}l@{}}Ensure\\ Privacy\\ Practices\end{tabular}                                                      & \fivepointlikertt{3}{2}{2}{1}{0}                                      &                     &                                                                                  &                                                                      \\
\cellcolor[HTML]{EFEFEF}\begin{tabular}[c]{@{}l@{}}Challenges in\\ Anticipating\\ Contributor\\ Attrition\end{tabular}        & \cellcolor[HTML]{EFEFEF}\fivepointlikertt{3}{2}{2}{1}{0}              &                     & \cellcolor[HTML]{EFEFEF}\begin{tabular}[c]{@{}l@{}}Automated\\ Notifications \end{tabular} & \cellcolor[HTML]{EFEFEF}\fivepointlikertt{5}{2}{1}{0}{0}               &                     & \cellcolor[HTML]{EFEFEF}                                                         & \cellcolor[HTML]{EFEFEF}                                             \\
\begin{tabular}[c]{@{}l@{}}Challenges in\\ Ensuring Data Privacy\\ While Tracking\end{tabular}                                & \fivepointlikertt{3}{1}{3}{1}{0}                                       & \multirow{-10}{*}{} & \begin{tabular}[c]{@{}l@{}}Utilize Tools for\\ Contributor\\ Engagement\end{tabular}                                      & \fivepointlikertt{5}{1}{2}{0}{0}                                       & \multirow{-10}{*}{} &                                                                                  &                                                                      \\
\multicolumn{8}{l}{\cellcolor[HTML]{EFEFEF}\begin{tabular}[c]{@{}l@{}}The dark green section represents the percentage of participants who “Strongly Agree,” while the light green section shows those who “Agree.”\\ The gray section corresponds to “Neutral,” the light red corresponds to “Disagree,” and the red section to “Strongly Disagree.\end{tabular}}                                                                                                                                                                                                                                      \end{tabular}
}
\label{tab:user_study}
\end{table}

%% file: section/sec8_Discussion.tex
\section{Discussion}

As we observed, managing contributor retention in OSS is a multifaceted challenge, requiring a balance between automated mechanisms, privacy concerns, and access to retention-related data. Our findings highlight several tensions that emerge in attempting to design effective retention strategies while ensuring inclusivity and fairness.

\noindent\textbf{Privacy vs. Data Aggregation.} One tension we observed lies in the balance between protecting contributors' privacy and leveraging data aggregation for retention insights. While we can use automated data collection to track contributors' activities and predict attrition, it also raises ethical concerns regarding the surveillance of these contributors. 
Our findings indicate that while community managers benefit from tracking contributor activity to anticipate disengagement (C9), there are significant concerns regarding data privacy and potential misuse (C10). Some participants hesitated to implement predictive models due to ethical concerns surrounding monitoring contributor behavior. The aggregation of contributor activity may reveal individual disengagement patterns, potentially impacting personal autonomy and raising concerns about data ownership.

\noindent\textbf{Automated Notifications and Engagement Fatigue.} Another point that was evidenced during our study was the role of automated notifications and outreach in managing retention. While the literature showed that automated systems can help community managers reduce their workload (S8) \cite{wessel2021don}, several participants (P2, P4, P6) reported concerns that frequent notifications might contribute to notification fatigue and disengagement (C2). Participants noted that contributors who receive excessive automated messages may feel overwhelmed, perceiving a lack of personal touch in interactions. Our results also suggest that the impact of automated outreach varies depending on project size and governance structures. That said, it becomes important to work on further understanding how personalization and contextual adaptation may be used in each specific case to avoid harming the project instead of helping the maintainers.

\noindent\textbf{Cross-sectional vs. Longitudinal Retention Insights.} One interesting point to discuss regards whether cross-sectional (snapshot-based) or longitudinal (historical) approaches provide more useful insights for retention tracking. Our study found that many OSS projects rely on cross-sectional retention analyses due to time and resource constraints (C5, C6). However, this approach often fails to capture long-term contributor trajectories, leading to misinterpretations of short-term fluctuations as retention failures. On the other hand, longitudinal tracking offers more profound insights into contributor retention trends, but it also introduces complexity and noise (C1, C2). OSS community managers indicated that while they value historical retention data, sustained monitoring is challenging due to competing responsibilities (P2, P3, P6). This underscores the need for hybrid approaches that integrate both short-term and long-term retention signals, allowing maintainers to identify both immediate risks and broader systemic trends.

\input{table/trigu}

\noindent\textbf{Triangulation of Our Findings.} We used triangulation to ensure the robustness and reliability of our findings: challenges in managing retention and strategies in supporting contributor retention. The summary of our triangulation is provided in Table \ref{tab:trigu}.

We first investigated the challenges of managing retention in OSS through multiple triangulation approaches. We began by synthesizing challenges from interviews with experienced OSS managers whose extensive expertise laid the foundation for understanding these challenges. We then triangulated these findings through a multi-vocal literature review of academic and community-driven gray literature. Finally, we surveyed with OSS community managers with varying experience levels as a final validity check. Refer back to Table \ref{tab:monitor_challenges} for a summary of the complete triangulation process and its results.

As for strategies in supporting OSS community managers in diagnosing and managing retention challenges, to ensure their helpfulness and practicality, we also employed multiple triangulation validation strategies. We began by synthesizing strategies through literature reviews. We then operationalized them in a web-based prototype and conducted an in situ evaluation. The prototype was designed using a user-centric approach, engaging over 100 OSS community managers throughout the iterative development process to ensure the operationalized strategies in our prototype were aligned with real-world needs. We then conducted an in situ evaluation within two OSS communities. These evaluations confirmed their helpfulness and generalizability to real-world OSS management. The complete results are summarized in Table \ref{tab:strategies}.

\noindent\textbf{Impact of Our Work.} Our work addresses challenges in managing OSS communities, particularly in managing contributor retention, which has become a distinct, complex task \cite{gray2022disengage}. In larger communities, this responsibility often requires dedicated teams, reflecting the growing importance of contributor management \cite{gray2022disengage, zhou2016inflow}. Our research offers a consolidated list of challenges, enabling future research and OSS governance committees to understand these issues better and address them effectively. Meanwhile, our work also highlights the importance of responsibility in managing contributors' retention. Such emphasis helps OSS governance committees become aware of and acknowledge community managers' responsibilities and difficulties faced.

We provided a set of strategies to support OSS community managers in diagnosing and responding to contributor retention challenges.  These strategies were integrated into a prototype that demonstrates how they can be operationalized in practice. Validated through multiple rounds of stakeholder discussions and in situ evaluations, the strategies provide a guidebook that OSS communities can adapt to their specific needs. For example, current tools and practices in OSS governance could integrate the strategies we synthesized to enhance their design and better address the challenges faced by community managers.

Furthermore, our research approach serves as a showcase for identifying and establishing the ground truth of challenges and strategies to OSS ecosystems' unique social and organizational dynamics. Our design approach is rooted in human-centered practices. By engaging stakeholders throughout the process, we ensured that the strategies we reported in this study aligned with their needs, goals, and challenges.

\noindent\textbf{Future Design Opportunities: Incorporating Design Strategies with Large Language Models:} Given the increasing prevalence of large language models (LLMs) and generative AI \cite{hacker2023regulating}, integrating these technologies with our design strategies can help automate retention management tasks in OSS communities (S8 and S9). LLMs can enhance automated notifications by generating personalized project activity summaries and identifying signs of contributor disengagement. By analyzing contributor activity patterns, LLM-powered alerts can provide community managers with actionable insights, such as flagging contributors who may need additional support or recognizing emerging challenges that impact project health. Additionally, LLMs can generate contextualized messages that balance encouragement and information, reducing the likelihood of generic or overwhelming notifications. Moreover, LLMs can improve contributor engagement by personalizing engagement strategies. For example, community managers may not even need to draft a generic onboarding message. LLMs can tailor recommendations for tasks, learning resources, and mentorship opportunities based on contributors' prior activity and skill levels.

\noindent\textbf{Not All Turnover Is Equal: Rethinking Retention Metrics.} While our study focuses on surfacing and responding to retention challenges, it is essential to recognize that not all contributor departures are inherently negative. Departures may occur for various reasons, some of which reflect natural or beneficial lifecycle events. For instance, contributors may disengage after completing a specific feature they prioritized \cite{barcomb2019retaining};  a pattern observed in our findings (C3, P5). Similarly, maintainers may step down after years of service to avoid burnout and enable new leadership, fostering community renewal \cite{miller2019people, lin2017developer, Zhou2010DeveloperFA}. Such transitions signal healthy community dynamics and should not be treated as retention failures.

By contrast, departures driven by exclusion, burnout, or unresolved interpersonal conflicts pose significant risks to community sustainability \cite{miller2019people}. Future research could explore a typology of departures, such that they are categorized as concerning (e.g., exclusion-driven), natural (e.g., task completion), or positive (e.g., leadership transitions), to guide context-sensitive retention strategies. Such a framework could enhance diagnostic tools, enabling community managers to prioritize interventions to avoid negative disengagement while acknowledging natural or positive transitions. Longitudinal studies of contributor trajectories could validate this typology by identifying patterns of exit intent and informing more effective retention practices, as they could observe changes over time to trace natural contributor behavior patterns \cite{easterbrook2008selecting}.

%% file: table/trigu.tex
\begin{table}[!tbp]
\caption{Triangulation of Challenges and Strategies in Managing Retention in OSS}
\resizebox{5.5in}{!}{
\begin{tabular}{lccccc}
\rowcolor[HTML]{EFEFEF} 
\multicolumn{1}{c}{\cellcolor[HTML]{EFEFEF}} & \textbf{\begin{tabular}[c]{@{}c@{}}Interview\\ \faComments\end{tabular}} & \textbf{\begin{tabular}[c]{@{}c@{}}Survey\\ \faPollH\end{tabular}} & \textbf{\begin{tabular}[c]{@{}c@{}}Literature Review\\ \faBook\end{tabular}} & \textbf{\begin{tabular}[c]{@{}c@{}}User Centric Design\\ \faUsers\end{tabular}} & \textbf{\begin{tabular}[c]{@{}c@{}}In-situ User Evaluation\\ \faUserEdit\end{tabular}} \\
\textbf{Challenges}                          & \checkmark                                                               & \checkmark                                                         & \checkmark                                                                   &                                                                                  &                                                                           \\
\rowcolor[HTML]{EFEFEF} 
\textbf{Strategies}                          &                                                                          &                                                                    & \checkmark                                                                   & \checkmark                                                                       & \checkmark                                                               
\end{tabular}}
\label{tab:trigu}
\end{table}

%% file: section/sec9_Threats.tex
\section{Limitations}

Like any empirical study, our research contains limitations. Here, we discuss these limitations and the measures we employed to address them.

The initial list of challenges was derived from the six participants, which may lead some to question whether the challenges we identified were representative enough. To account for this, we analyzed for saturation, and no new challenges emerged in the last two interviews.  In qualitative analysis, the adequacy of sample size is determined by whether data saturation is reached, rather than a minimum number of interviews \cite{sebele2020saturation, saunders2018saturation, caine2016local}. Although the number of participants required to reach saturation is debated \cite{townsend2013saturation}, prior empirical work suggests that many core themes emerge within the first six interviews \cite{guest2006many}. Some qualitative approaches also use small samples (3–10 participants) \cite{creswell2016qualitative}. To overcome any limitation from the number of interview participants, we triangulated the challenges by conducting multi-vocal literature reviews and expert evaluation surveys.  Our participants represented diverse backgrounds, including differences in gender, experience, and seniority in managing OSS communities.

Following the same methodology, the identified strategies were also triangulated from a literature review, a user-centric design approach, and in situ user evaluation involving over 100 stakeholders from OSS communities. This approach allowed us to gather rich, contextual insights and iteratively refine our strategies in real-world settings.  We operationalized our strategy into a prototype through user-centric design. This prototype represents one of countless possible ways to implement the identified strategies. While it aims to understand the helpfulness and practicality of these strategies, it is not intended as a fully scaled or universally applicable solution. Instead, the prototype serves as a proxy in our triangulation methodology to validate the practicality and helpfulness of the strategies in real-world OSS contexts.

During the prototype development phase, we adopted a user-centric design approach and iteratively collected feedback through three rounds of focus group discussions from different conferences across different communities and regions. Due to conference constraints, we did not collect detailed demographic data from focus group participants. However, to enhance generalizability, we conducted these sessions across multiple OSS venues (e.g., FOSSY, Linux Plumbers, OpenUK) and engaged participants from a range of communities and geographic regions.

To evaluate strategies and ground our findings in authentic usage settings, we conducted an in situ user evaluation with managers from two OSS projects to assess the operationalized strategies embedded within our web-based prototype. While such an evaluation does not capture long-term behavioral adoption or usage trajectories, it does allow managers to engage with strategy representations using their own project data and decision-making processes, yielding context-rich and temporally grounded insights \cite{consolvo2007conducting, lechelt2021evalme}. Moreover, we opted not to conduct a controlled study, as it was not deemed necessary for our research; the goal of the evaluation was to understand the perceived helpfulness of the strategies rather than to compare the tool itself against another tool \cite{gooch2016creating}.

Another concern could be the representativeness of the number of evaluation participants. In addition to meeting the qualitative analysis requirements mentioned in the first limitation, this user evaluation is just one step in our triangulation process, and the overall exploration of strategies has involved over 100 stakeholders. Moreover, our user evaluation was conducted across two communities with varying natures and scopes, as discussed in Section \ref{sec:evaluation}, allowing us to capture nuanced perspectives on how these strategies function in practice.

Finally, regarding the predictive models for forecasting contributor attrition, different models exhibit varying levels of accuracy. Therefore, we provided a set of three models along with the accuracy information of each model to help users select the model that is most appropriate for their needs.

%% file: section/sec10_Conclusion.tex
\section{conclusion}

In this paper, we report ten challenges and nine strategies for managing contributor retention in OSS. Our findings are grounded in multi-triangulation, with over 100 stakeholders participating in our project. Our results highlight both the complex responsibilities community managers face in supporting contributor retention, as well as a set of strategies to inform governance decisions and strengthen retention practices across OSS communities.

We believe that our study will be valuable to OSS communities by identifying challenges from multiple perspectives and equipping community managers with data-driven strategies to foster sustainable engagement. By bridging the gap between academic research and real-world OSS practices, our work showcases a human-factors-oriented framework that serves as a foundation for future researchers developing theories on OSS sustainability. Looking ahead, our future work will expand on this foundation by scaling the strategies and exploring how AI can enhance retention efforts.

The research artifacts for this study are available publicly at the
companion website \cite{suppl}.

\section*{Acknowledgment}

We thank all participants in our study, including those who contributed through interviews, surveys, and design discussions, for their time and valuable insights. We are especially grateful to the managing teams of the Pyomo and DeepSpeed projects for their participation and support. We also thank Edward Gilmour for his helpful feedback during the preparation of this paper. This work was partially supported by NSF Grant Nos. 2303043, 2303042, and 2303612.